\documentclass[12pt,a4paper]{iopart}

\usepackage{rotating,graphicx}
\usepackage{amsfonts}%
\usepackage{caption}
\usepackage[utf8]{inputenc}
\usepackage[english]{babel} 
\usepackage[T1]{fontenc}
\usepackage{color}
\usepackage{setspace}
\usepackage{layouts}
\begin{document}

\title[On the role of filaments in SOL heat transport]{On the role of filaments in perpendicular heat transport at the Scrape-off Layer}
\author{D. Carralero $^{1,2}$, S. Artene$^{1}$, M. Bernert$^{1}$, G. Birkenmeier$^{1}$, M. Faitsch$^{1}$, P. Manz $^{1,3}$, P. deMarne$^{1}$, U. Stroth$^{1}$, M. Wischmeier$^{1}$, E. Wolfrum$^{1}$, the ASDEX Upgrade team$^{4}$ and the EURO-fusion MST1 Team$^{5}$.}
\address{$^1$ Max-Planck-Institut für Plasmaphysik, 85748 Garching, Germany.}
\address{$^2$ Laboratorio Nacional de Fusión, CIEMAT, Av. Complutense 40, 28040 Madrid, Spain}
\address{$^3$ Physik-Department E28, Technische Universität München, Garching, Germany}
\address{$^4$ See the author list A. Kallenbach et al. Nucl. Fusion 57 (2017) 102015}
\address{$^5$ See the author list H. Meyer et al. Nucl. Fusion 57 (2017) 102014}
\ead{daniel.carralero@ipp.mpg.de}

\begin{abstract}

In this work we carry out quantitative measurements of particle and heat transport associated to SOL filaments in a tokamak, and relate density shoulder formation to the advection of energy in the far SOL. For the first time, this attempt includes direct measurements of ion and electron temperatures for background and filaments. With this aim, we combine data from a number of equivalent L-mode discharges from the ASDEX Upgrade tokamak in which different probe heads were installed on the midplane manipulator. This approach is validated by a  comparison with independent diagnostics. Results indicate an increase of heat transport associated to filaments after the shoulder formation. Several centimeters into the SOL, filaments are still found to carry a substantial fraction (up to one fifth) of the power ejected at the separatrix.

\end{abstract}

\maketitle

\section{Introduction}\label{intro}

Perpendicular transport in the Scrape-off Layer (SOL) of confined fusion devices is generally accepted to be dominated by elongated convective cells known in the literature as filaments or blobs \cite{Endler95,Dippolito11}. In particular, there is plenty of experimental evidence that perpendicular particle transport in tokamaks is substantially enhanced by these structures: from the onset of the ``main chamber recycling regime'' in Alcator C-mod \cite{Umansky98} to recent experiments in ASDEX Upgrade (AUG) \cite{Carralero14}, this has been observed in most major machines, such as JT-60U \cite{Asakura97}, DIII-D \cite{Boedo01}, TCV \cite{Garcia06} or MAST \cite{Kirk16}. Some of these studies have been related to the flattening of the SOL density profile known as ``shoulder formation'' and observed in many tokamaks after a certain density is achieved in L-mode operation, typically in the range of 0.45 times the Greenwald density limit \cite{Greenwald02}. The onset of the shoulder has been explained by models as the result of the electrical disconnection of filaments from the wall caused by the increasing collisionality \cite{Myra06}. According to this, filaments would change their propagation regime, increasing their radial velocity and size, thus enhancing perpendicular transport and causing the shoulder. In previous works, evidence was found of such a mechanism in the AUG and JET tokamaks \cite{Carralero14, Carralero15}, where the flattening of the density profile could be linked to the increase of collisionality in the divertor region and to the change in the filament properties. Later studies have tried to extend these results into H-mode \cite{Carralero16, Wynn18} and tokamaks with different geometries \cite{Vianello17} finding that increased collisionality is not a sufficient condition for the shoulder formation, and that neutrals also play a necessary role in this phenomenon. In a recent work, a general framework has been proposed in which EMC3-EIRENE simulations and experimental data of the shoulder formation in AUG are explained by a combination of enhanced filamentary transport and decreased neutral mean free path at the midplane which acts as a fueling barrier, bringing the ionization front towards the SOL \cite{Carralero17}.\\

The discussion on filaments and their impact on perpendicular transport in the SOL has nevertheless been mostly focused on particle transport, but very few experimental works have tried to measure their impact on perpendicular heat fluxes and energy transport. Even in the cases were this was achieved, only the electron contribution has been taken into account (eg., in Alcator C-mod \cite{LaBombard01} or DIII-D \cite{Boedo09}) or broad estimations were required in order to include ions (eg., in AUG \cite{Birkenmeier15}). The main reason for this is instrumental: while several common SOL diagnostics can measure density or density-based fluctuations on the required time scales (such as probes, cameras, alkali beams, etc.), fast enough temperature measurements are difficult to realize in this region. This is especially the case for the ion temperature and even general models have until recently assumed cold ions for simplicity \cite{Krash08}. In recent years, some models have explored the role of $T_i$ on filament propagation \cite{Manz13} , and several experiments have been specifically aimed to measure the ion temperature of filaments, $T_{i,fil}$, both in AUG \cite{Kocan12} and MAST \cite{Allan16}. Also, the evolution of $T_i$ throughout the shoulder formation has been studied in both tokamaks \cite{Elmore12,Carralero17}. However, no direct measurement of the total heat flux (ie., including the contributions of ions and electrons) associated to filaments has been conducted so far. This question is of obvious importance for future reactors, as the transport of energy in the SOL will determine both the width of heat loads onto the divertor targets and the sputtering levels on the plasma facing components (PFC) at the main wall. Therefore, the first objective of the present work is to find an experimental answer to the general question ``how much energy do filaments transport?''. Also, since particle transport towards the first wall is substantially enhanced with shoulder formation   \cite{Carralero14, Guillemaut17}, it is only natural to wonder if it has a similar impact on the heat fluxes. Therefore, as a second objective we also set out to answer the question ``Is the particle transport increase observed during the shoulder formation followed by an equivalent increase of energy transport?''. \\

In order to answer these questions, a comprehensive set of relevant and reproducible L-mode discharges from AUG has been taken, such that, by combining the different measurements carried out, all the necessary information can be put together in order to estimate the heat transport in the outer midplane SOL (namely, density, ion and electron temperatures, radial velocity and filament packing fraction) for typical SOL conditions before and after the shoulder formation. Of course, all the necessary diagnostics were never available at the same time: paradigmatically, the midplane manipulator (MPM) was not able to measure filamentary properties and $T_i$ at the same time, since different probe heads were required (namely a pin probe and a retarding field analyzer). In spite of this and the fact that some of the experimental data had to be interpolated between different discharges, this analysis should  provide an reasonable quantitative answer to the question. Besides, the excellent agreement with independent diagnostics (such as infrared cameras and lithium beam) indicates that the approximation is sound and the results realistic. This work is a continuation of the AUG studies discussed previously \cite{Carralero14,Carralero15,Carralero16,Carralero17}, and most of the details regarding the experimental setup and analysis techniques are already described there.\\

This paper is organized as follows: First, in section \ref{Exp}, the experimental  database is described, along with the diagnostics involved. In section \ref{Mes} the SOL measurements combined in the database are displayed. In section \ref{sec:validation}, these combined measurements are validated by comparing them to measurements from independent diagnostics. Next, in section \ref{sec:transport}, perpendicular particle and heat transport associated to filaments are calculated and discussed. Finally, conclusions are drawn in section \ref{Con}.

\section{Experimental Setup}\label{Exp}

As discussed in the introduction, recent work carried out on AUG shows that the divertor collisionality, $\Lambda_{div}$, is the control parameter for the L-mode filamentary transition and the shoulder formation. Therefore, this parameter is used to classify the data according to the degree of shoulder formation. $\Lambda_{div}$ is defined as in \cite{Carralero15}:
\begin{equation}
\Lambda_{div}=\frac{L_\parallel/c_s}{1/\nu_{ei}}\frac{\Omega_i}{\Omega_e}, \label{eq2}
\end{equation}
 where $\nu_{ei}$ is the electron-ion collision rate, $c_s$ is the sound speed and $\Omega_{i/e}$ stands for the gyrofrequency of ions/electrons. $L_\parallel$ is the characteristic parallel length in the divertor region, which is defined as $1/5$ of the connection length between the midplane and the closest divertor target (placed in the LFS). As explained in \cite{Carralero15}, the factor $1/5$ gives an estimate of the length of the field line below the X-point. According to this definition, disconnection takes place for $\Lambda>1$, when the characteristic parallel transport time is longer than the inverse of ion-electron collision frequency. For each data point, $\Lambda_{div}$ is calculated using time-averaged $n_e$ and $T_e$ values measured by divertor Langmuir probes in the $\rho_p \in [1.02,1.03]$ region of the divertor (see Fig. \ref{fig:div} for typical profiles), where $\rho_p = \sqrt{\Psi_p/\Psi_p^{sep}}$ is the radial magnetic coordinate defined by the enclosed poloidal magnetic flux, $\Psi_p$, normalized by its value at the separatrix. Three ranges of $\Lambda_{div}$ values are defined to order the data: low collisionality ($\Lambda_{div} < 0.3$), for which no shoulder has yet formed,  high collisionality ($\Lambda_{div} > 3$), for which the shoulder is already formed, and medium collisionality ($0.3 < \Lambda_{div} < 3$), which lies on the transition between the two previous situations. These three ranges, represented in figures respectively with red/blue/black colors, will be used throughout the work to characterize the data.\\

\begin{figure}
	\centering
		\includegraphics[width=0.5\linewidth]{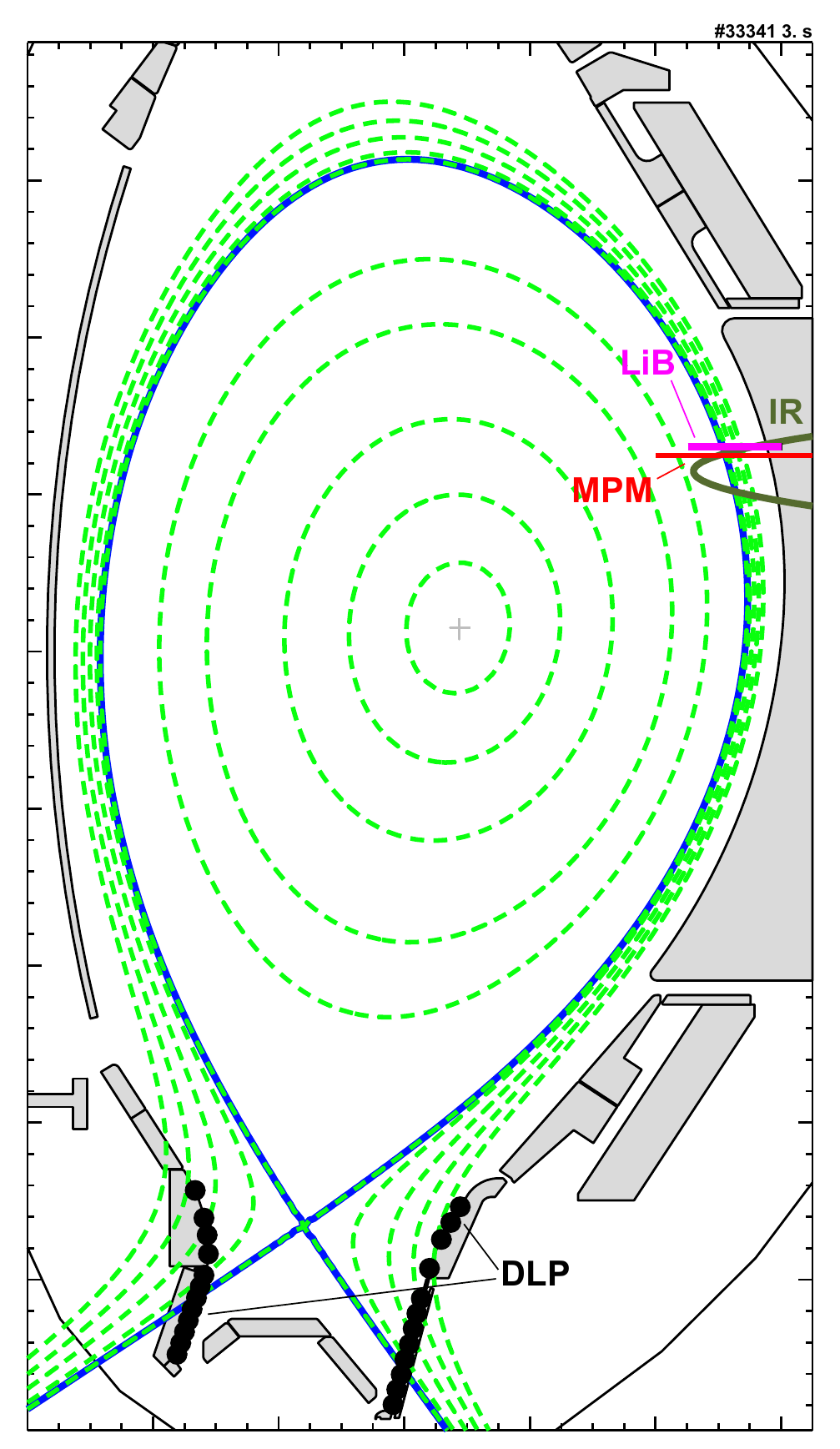}
	\caption{\textit{Poloidal view of the AUG plasma during a typical discharge. Separatrix and other flux surfaces are represented as solid blue and dashed green lines. The position of the main diagnostics employed in this work are indicated, including the line of sight of the IR camera.}}
	\label{fig:diag}
\end{figure}

The experiments reported in the present work were carried out in AUG, a medium-sized divertor tokamak with full tungsten coated walls \cite{Kallenbach11}. Its major and minor radii are $R = 1.65$ m and $a = 0.5$ m, respectively. All discharges consisted of lower single-null L-mode plasmas with common magnetic parameters (toroidal magnetic field $B_t = -2.5$ T, edge safety factor $q_{95} = 4.9-5.3$, plasma current $I_p = 800$ kA) and an edge optimized configuration, featuring flux surfaces roughly parallel to the outer limiter. A typical poloidal view of this configuration can be seen in Fig. \ref{fig:diag}. Some discharges (such as the ones described in \cite{Carralero17}) consisted of density ramps in which the collisionality threshold was crossed. Different heating powers were used, ranging from purely ohmic discharges to up to $600$ kW of ECH power. Others (like the ones described in \cite{Carralero14}) consisted in constant-density discharges in which the radial position of the separatrix was varied in order to measure profiles. In total, around 60 data points were collected in 20 discharges carried out in the 2013, 2014 and 2016 AUG campaigns.\\

The layout of the main diagnostics used in this work is displayed in Fig. \ref{fig:diag}. The midplane manipulator (MPM) and the divertor Langmuir probes (DLP) have been used to form the database from which the SOL transport was calculated. Instead, as will be described in Section \ref{sec:validation}, the lithium beam (LiB) and an infrared (IR) camera overlooking the MPM were used to benchmark the results obtained from the database.\\

The MPM is inserted horizontally into the LFS of the plasma around 30 cm above the midplane and provides measurements of profiles and turbulence. It was equipped with three different probe heads including: a) A multi-pin probe (named ``14-pin probe'' and described in detail in \cite{Nold12}), which was used to measure ion saturation current ($I_{sat}$) in several Langmuir pins separated radially and poloidally. Part of the discharges also featured a sweeping pin, providing measurements of $T_e$ \cite{Hutchinson}, and some featured a Mach probe, yielding measurements of the parallel Mach number, $M_{||}$ \cite{Hutchinson88}. b) A Retarding Field Analyzer (RFA), used to obtain measurements of $T_i$. This probe head is described in detail in \cite{Kocan12}. c) A high heat flux pin probe (HHFP), which was developed in 2016 in order to carry out measurements equivalent to those of the 14-pin probe in the vicinity of the separatrix.\\

The main purpose of the MPM is the evaluation of filamentary transport in the midplane. Following the analysis techniques described in previous work \cite{Carralero14}, pin probes with high temporal resolution DAQ systems (sampling at 2 MHz) are used to detect filaments (defined as events in which the $I_{sat}$ signal at the reference pin exceeds $2.5$ times the standard deviation $\sigma$) and to measure several turbulent characteristics such as the auto-correlation time $\tau_{AC}$, filament detection rate, etc. These measurements are then used to calculate the radial and binormal velocities of the average filament, as well as the packing fraction of the filaments. Sweeping pins are also used to estimate the filament and background $T_e$ separately: using the $I_{sat}$ signal from the reference pin to define filaments, both the collected current and the corresponding polarization values at the sweeping pin can be conditionally averaged. By these means, two I-V curves can be obtained for filaments and background, on which different $T_{e,fil}$ and $T_{e,back}$ values can be fit.\\

A similar approach is used with the RFA probe: this diagnostic collects ions entering a cavity with a grid at the entrance. Outside the cavity, a plate with a slit is polarized to a negative potential in order to deflect incoming electrons (and also to obtain a measurement of the local $I_{sat}$ value). As explained in detail in \cite{KocanRSI}, if the potential of such a grid, $V_{grid}$, is swept with respect to the plasma, the energy of the ions entering the cavity can be selected. By doing so, the collected current, $I_{col}$ can be expressed as a function of the ion energy. Assuming a Maxwellian distribution in the ion population far from the probe and a dominant ion charge $eZ_i$ in the incoming flux, the ion temperature can be estimated from $I_{col}$,

\begin{equation}
I_{col} \simeq I_{col,0} \exp\biggl[-\frac{eZ_i}{T_i}(V_{grid}-|V_s|)\biggr],
\end{equation}
where $V_s$ is the sheath potential drop at the entrance of the slit plate. Again, $I_{sat}$ values measured at the slit plate can be used to tell apart times when the $I_{col}$ measurement can be associated to a filament or to background plasma levels, thus obtaining two different $I_{col}$,$V_{grid}$ curves from which $T_{i,fil}$ and $T_{i,back}$ values can be fit. This technique was used in the past to measure $T_i$ associated to ELM events in \cite{Kocan12ELM}.\\ 

Density and temperature measurements from the divertor are carried out using the fixed flush-mounted Langmuir probes installed at the target plates (see Fig. \ref{fig:diag}). These probes are arranged in toroidal groups sharing the same poloidal position and connected as triple probes, thus yielding simultaneous measurements of $I_{sat}$ and $T_e$ \cite{Weinlich96}. At just 23 kHz, the sampling rate of divertor probes is substantially lower than that of the probe on the MPM, meaning that no turbulence analysis can be carried out from the data collected in them.\\

\begin{figure}
	\centering
		\includegraphics[origin=c,width=0.7\linewidth]{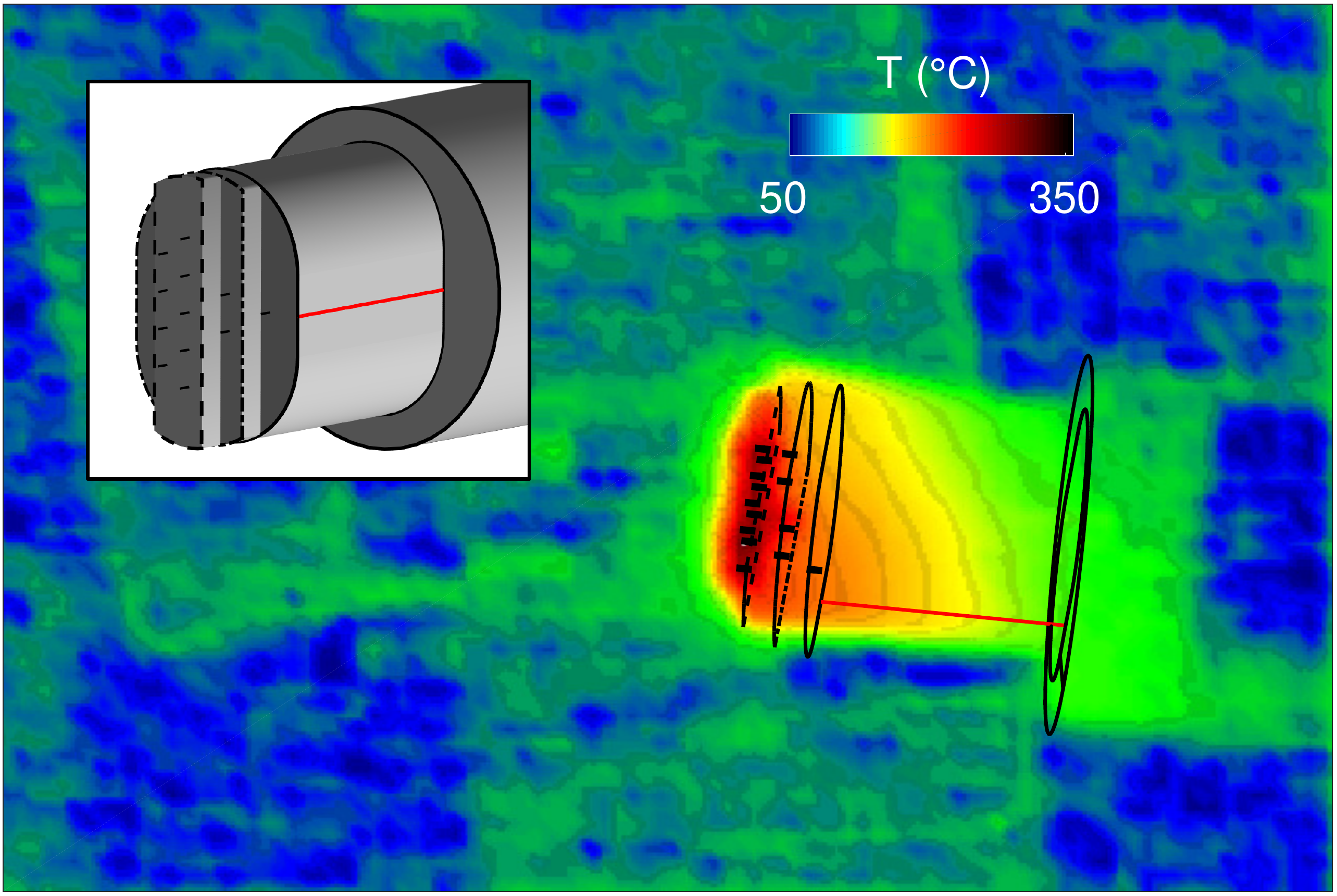}
	\caption{\textit{Typical IR thermography view of the 14-pin probe mounted on the MPM. Synthetic colors represent the surface temperature. In the insert, a 3D model of the 14-pin probe is displayed. In it, the edges of each terrace has been represented with different lines (dashed, dot-dashed, solid) along with the pins (short thick black lines) and a radial line along the probe shaft (red line). All these elements are projected on the thermography image.}}
	\label{fig:IR}
\end{figure}

Regarding diagnostics not included in the database, a lithium beam is used to measure density profiles in the SOL and edge of the plasma. As can be seen in Fig. \ref{fig:diag}, this diagnostic observes approximately the same poloidal region as the MPM (although both are separated by a toroidal angle of roughly 25$^{\circ}$). It features 26 radially distributed elliptical channels with a width of $6$ mm in the radial direction, covering the $\rho_p \in [0.9, 1.05]$ radial range \cite{Willendorfer13}. As well, in several discharges an IR camera observede the midplane region where the MPM was plunged, with the 14-pin probe mounted on it \cite{Carralero14}. After an absolute calibration was performed for a tungsten-covered body such as the 14-pin probe, this diagnostic measured the surface temperature of the probe during plunges. Once the jitter was removed, these data could be used to calculate the perpendicular heat flux on the probe surface using the THEODOR code \cite{Herrmann95,Herrmann01}. On Fig. \ref{fig:IR}, a typical thermography view of the 14-pin probe during a plunge is shown. In order to provide a meaningful interpretation of the data, a 3D model of the 14-pin probe is projected onto the IR picture. This model, shown in the insert of the figure, includes the perimeter of the three terraces where pins are mounted (displayed in the insert as dashed, dot-dashed and solid lines), along with the pins themselves and the edges of the surface at the end of the MPM shaft (also displayed as a solid black line). As can be seen in the figure, the projection of these curves fits nicely with the outline of the probe seen in the thermography view and the pins correspond to the hot spots detected in the upper terrace. In order to obtain a radial profile of the perpendicular heat flux onto the probe surface, the position of the region with perpendicular magnetic field incidence is also projected onto the image. As can be seen in the figure, this region (indicated as a red line) is also the one featuring the highest surface temperature for each radial position.\\

As will be discussed in section \ref{Temp}, several additional diagnostics have been used to complement probes in order to obtain temperature profiles. In the case of $T_e$, these include the edge Thomson scattering (TS) system and Electron Cyclotron Emission radiometer (ECE). The first includes 6 ND-YAG lasers at $1064$ mm, covering the edge of the plasma with scattering volumes close to the outer midplane and a repetition rate of 20 Hz \cite{Kurzan11}. The later involves an heterodyne radiometer which, at the magnetic configuration employed in this work ($B_t = -2.5$ T), covers the edge with 36 channels and a spatial resolution of about 5 mm. As well, a forward model for the electron cyclotron radiation transport is included \cite{Rathgeber13}. In order to combine the data from these two diagnostics at the edge, a Bayesian integrated data analysis (IDA) tool has also been employed \cite {Fischer10}. In the case of $T_i$, high resolution charge exchange recombination spectroscopy (CXRS) \cite{Viezzer12} was used to measure the temperature profile in the edge and near SOL of the plasma. In order to obtain data from this diagnostic, short NBI blips (in the order of 10 ms) have been used at the end of representative discharges. Combining toroidal and poloidal edge views of the system, characteristic $T_i$ profiles around the separatrix were obtained before and after the shoulder formation.\\

Finally, AUG bolometers \cite{Bernert14} are used to measure typical radiated power profiles for low and high density conditions. These measurements are used to estimate the power through the separatrix, $P_{SOL}$, 

\begin{equation}
P_{SOL}=P_{ohm}+P_{ECH}-P_{rad},
\end{equation}
where $P_{ohm}$ and $P_{ECH}$ are the ohmic and ECH heating powers, and $P_{rad}$ is the total power measured by bolometry radiated inside the separatrix.\\

\section{Measurements}\label{Mes}

\subsection{Midplane Fluctuations}\label{Fluc}

\begin{figure}
	\centering
		\includegraphics[width=.5\linewidth]{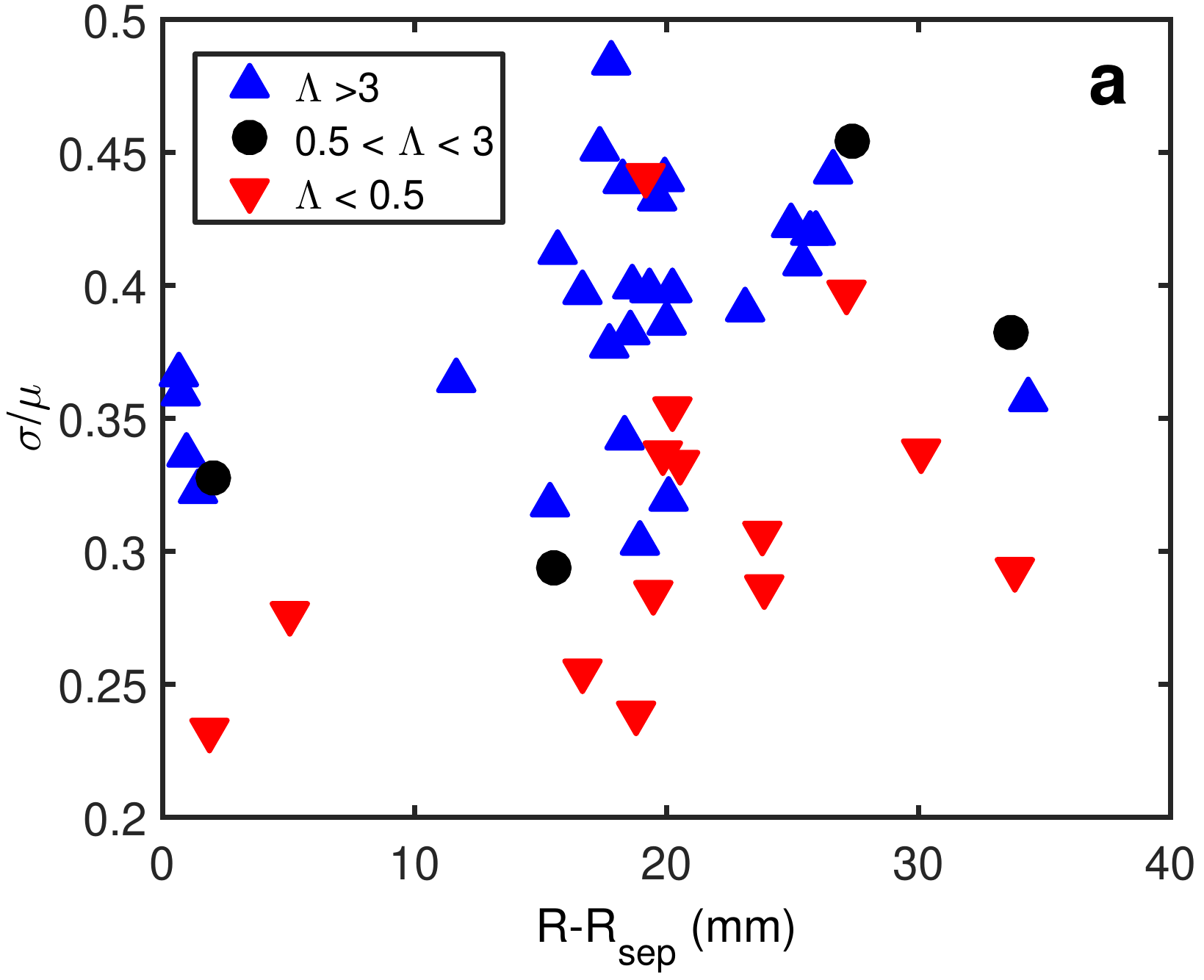}
		\includegraphics[width=.5\linewidth]{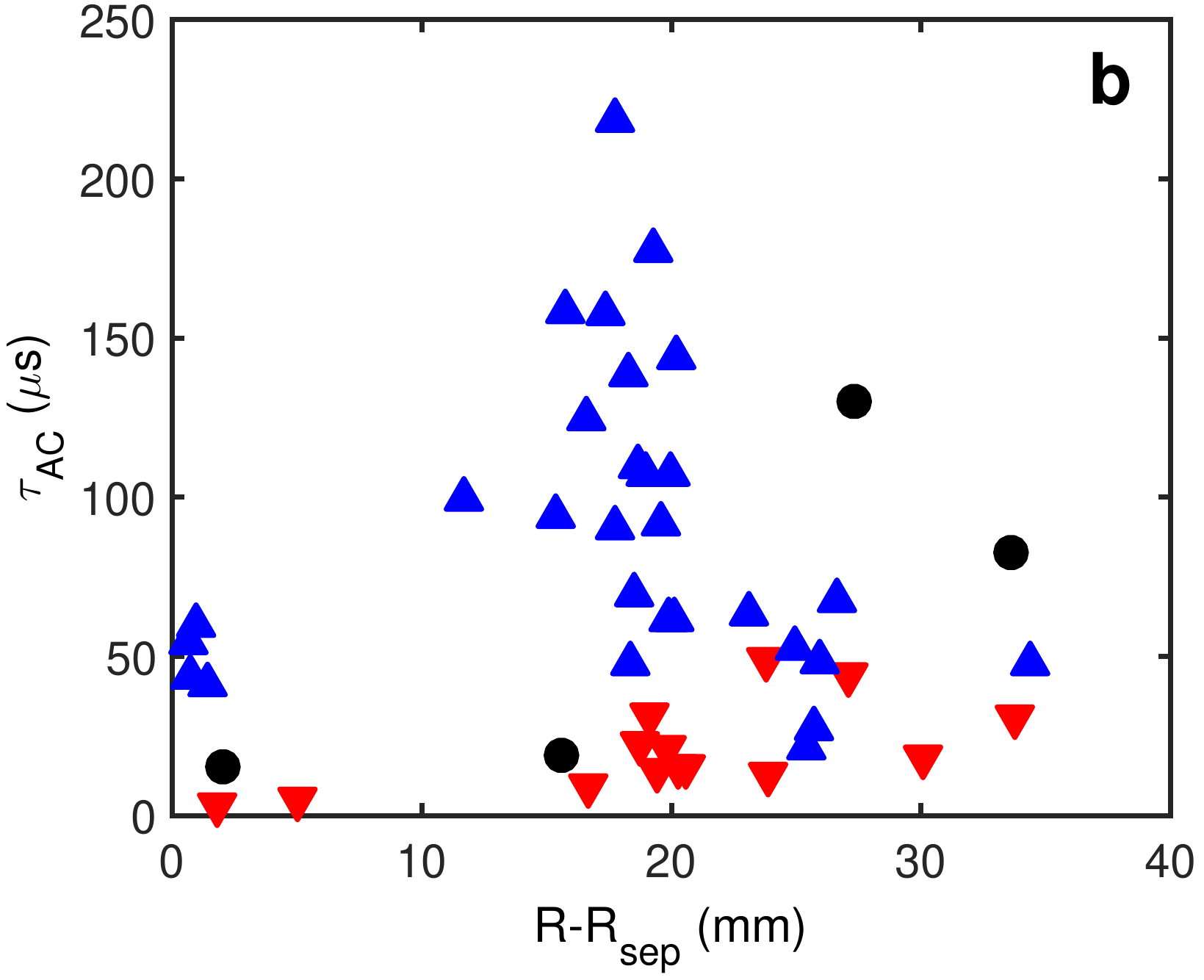}
		\includegraphics[width=.5\linewidth]{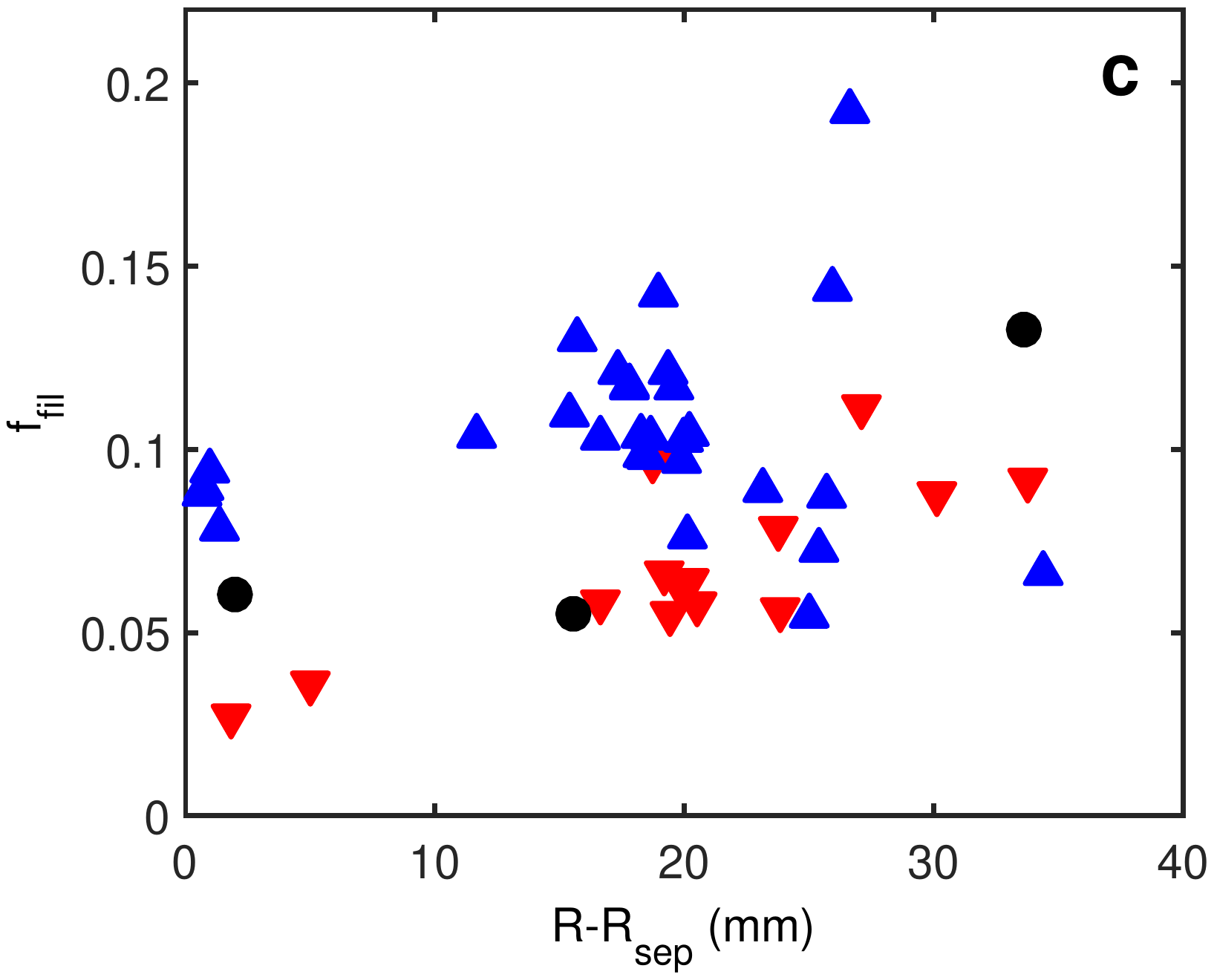}
	\caption{\textit{Turbulence property profiles derived from the 14-pin probe $I_{sat}$ measurements in the MPM. a) Relative amplitude of fluctuations, b) auto-correlation time and c) filament packing fraction, $f_{fil}$. Same color code as in Fig. \ref{fig:Turb}.} }
	\label{fig:Turb}
\end{figure}

As explained in the previous section, $I_{sat}$ measurements from the reference pin at the MPM were used to characterize turbulence in the outer midplane. In Fig. \ref{fig:Turb}, the profiles of three relevant parameters are displayed for the three collisionality ranges: First, in Fig. \ref{fig:Turb}a, the relative amplitude of the fluctuations is displayed. This is defined as the standard deviation of the fluctuations divided by the mean value, $\sigma/\mu$. As can be seen, the amplitude tends to grow with distance to the separatrix, and also seems to increase for intermediate and large $\Lambda_{div}$. In Fig. \ref{fig:Turb}b, the profiles of the auto-correlation time, $\tau_{AC}$ are presented. $\tau_{AC}$ is calculated using the standard definition found in literature (as the half width at half maximum of the auto-correlation function). In the first $20-25$ mm of the SOL, it can be seen how it is substantially increased after the shoulder begins to form, while further away from the separatrix the difference is not as clear. Both trends are consistent with previous observations of turbulence around the shoulder formation \cite{Carralero14}.

Fig. \ref{fig:Turb}c shows the filament packing fraction, $f_{fil}$, defined as

\begin{equation}
f_{fil}=\nu_{detect}\tau_{AC}, \label{eq4}
\end{equation}

where $\nu_{detect}$ is the filament detection frequency. This magnitude indicates the fraction of the total sample duration in which a filament is being detected by the probe. As with $\tau_{AC}$, $f_{fil}$ increases over a factor of 2 after the shoulder is formed in the first 25 mm of the SOL. By conditional analysis, the averaged properties of the filaments can be obtained for each data sample. In particular, using the relative delays of the filament detection times in several pins separated radially and poloidally, an estimation of the average radial and binormal velocity can be obtained (see the Appendix of \cite{Carralero17} for further details). Then, the average size of filaments in both directions can be obtained as the product of these velocities and the filament passage duration, $\tau_{AC}$. The evolution of filament size and velocity with $\Lambda_{div}$ was analyzed in previous work \cite{Carralero15} and is out of the scope of the present study.

\subsection{Midplane Temperature Profiles}\label{Temp}

As already explained in section \ref{Exp}, the main diagnostic for measuring temperature in the midplane was the MPM, either equipped with a sweeping pin in the case of $T_e$, or with an RFA in the case of $T_i$. In both cases, the amount of data is limited and MPM measurements had to be combined with additional diagnostics in order to include the near SOL. Even more, no profiles could be provided for the whole range of collisionalities discussed in this paper. Instead, a careful characterization was carried out for the low and high collisionality cases, and then the profiles corresponding to intermediate $\Lambda_{div}$ were interpolated between the two limit cases.\\

\begin{figure}
	\centering
		\includegraphics[width=.5\linewidth]{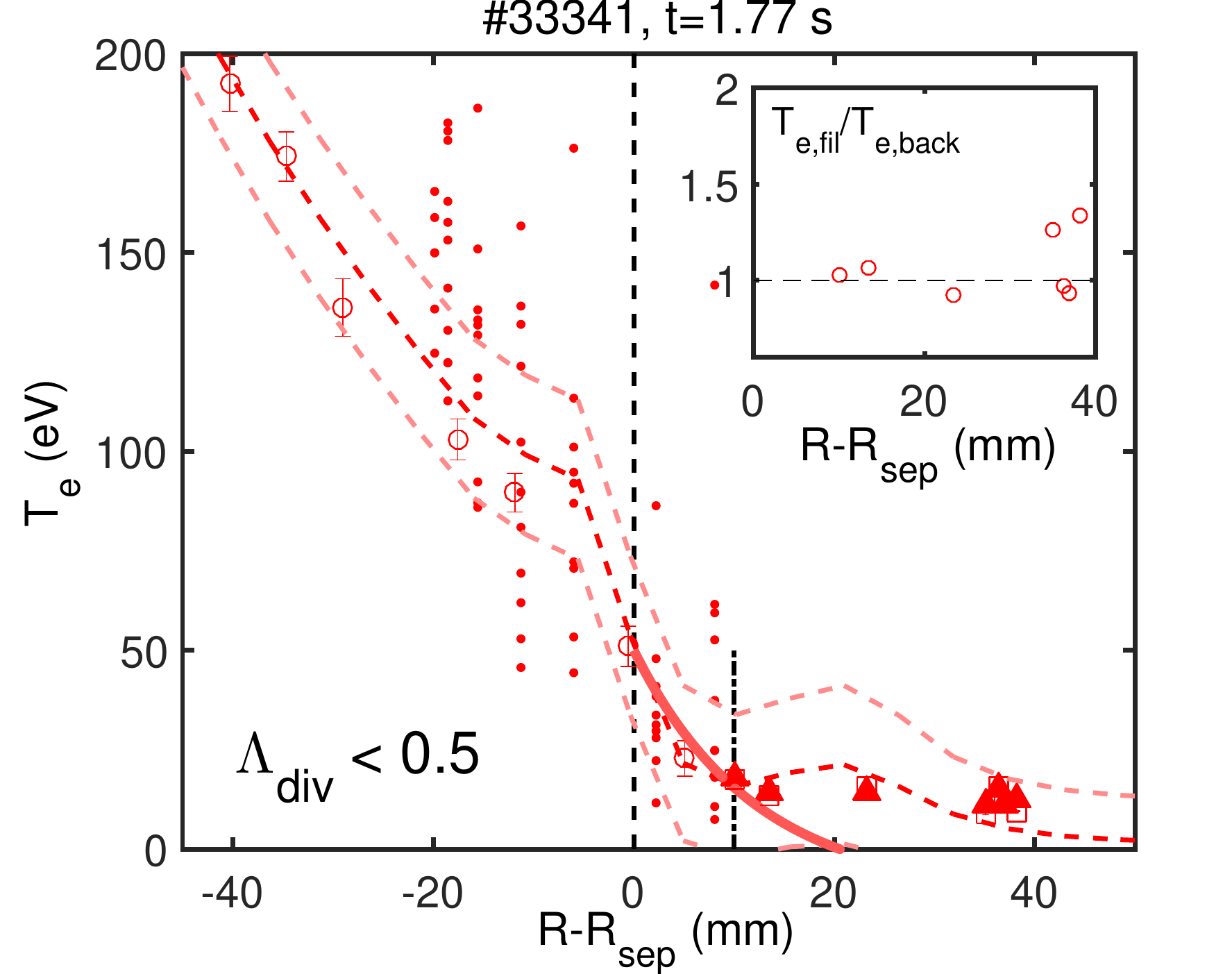}
		\includegraphics[width=.5\linewidth]{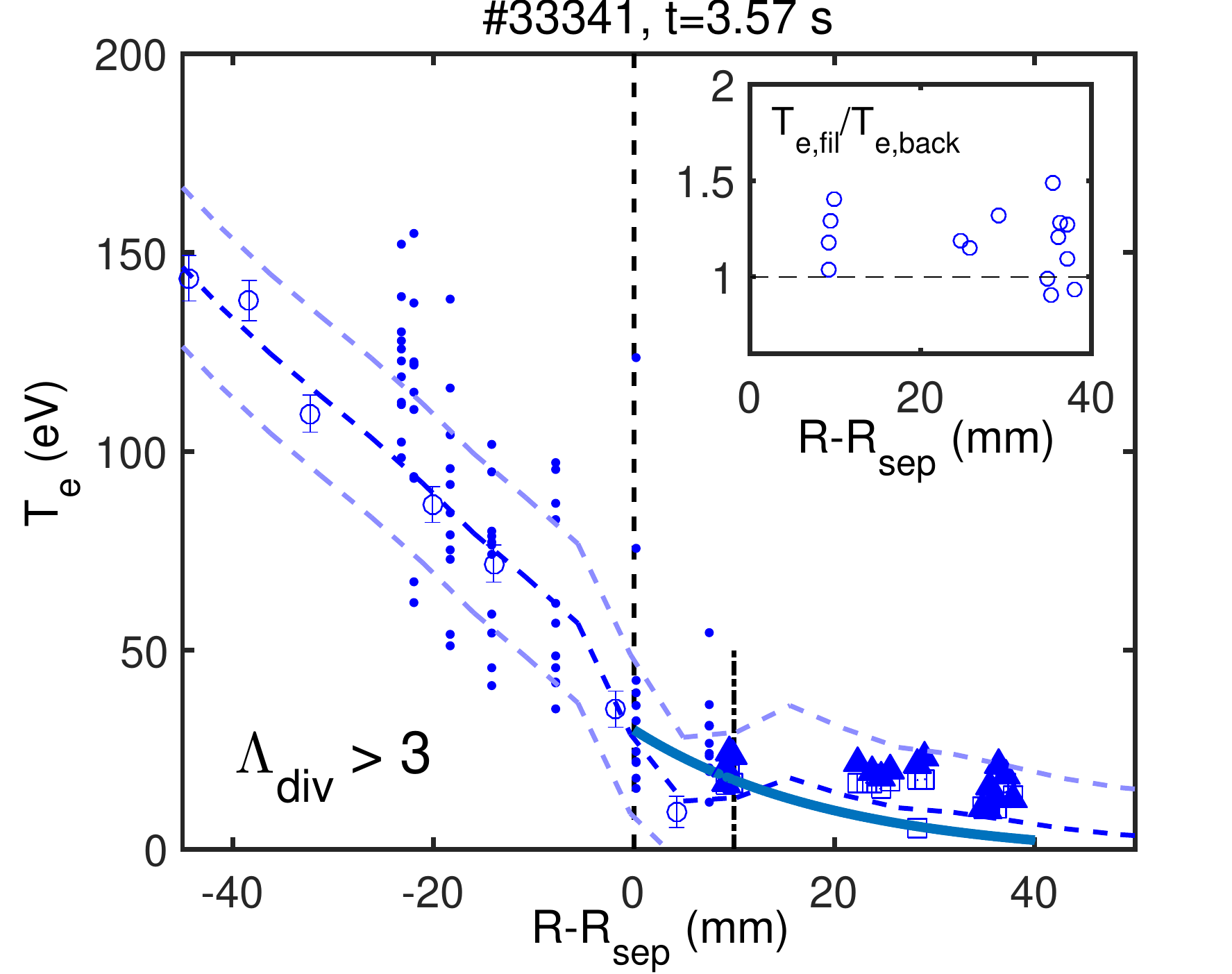}
	\caption{\textit{Electron temperature profiles for low/high $\Lambda_{div}$. Dots/circles represent TS/ECE data. Squares/triangles represent background/filament values as measured by the MPM manipulator. Dark/light dashed lines represent the mean value and the uncertainty of the edge IDA data. Thick solid lines represent the near SOL fit derived from the Thomson scattering $\lambda_{T_e}$. Vertical dash-dotted lines indicate $R-R_{SOL} = 10$ mm. In the inserts, the profiles of filament to background ratio $T_{e,fil}/T_{e,back}$ are displayed. Color code as in Fig. \ref{fig:Turb}}}
	\label{fig:Te}
\end{figure}

Characteristic low and high collisionality $T_e$ profiles are displayed in Fig. \ref{fig:Te}: Thomson scattering provides measurements in the region around the separatrix and ECE channels cover mostly the confined plasma edge. The measurements from both diagnostics are included in the IDA profile calculation (also displayed in the figure). However, neither of them is considered reliable for the far SOL, where the MPM is to be used instead. In order to define profiles for each case, an exponential curve is used in the near SOL, defined here as the first $10$ mm in front of the separatrix (the region between the separatrix and the dash-dotted line in the figure). The corresponding e-folding lengths are taken from TS data. According to recent L-mode measurements \cite{Sun17}, this means $\lambda_{T_e} \simeq 12$ and $\simeq 20$ mm for the low and high collisionality cases, respectively. As can be seen in the picture, this fit (represented by thick solid lines) provides a reasonable link between the IDA and the MPM data. For $R-R_{sep} > 10$ mm, typical probe measurements are taken instead. In order to define a different $T_e$ for filaments and background, only MPM data can be used. The $T_{e,fil}/T_{e,back}$ ratio, measured by the MPM as explained in section \ref{Exp} is displayed for each case in the inserts of Fig. \ref{fig:Te}. As can be seen, before the shoulder is formed, fluctuations seem to have a similar $T_e$ values as the background. Therefore, $T_{e,fil}\simeq T_{e,back}$ is assumed. For higher collisionalities, although data show some dispersion, filaments seem to have higher $T_e$. Therefore, taking an average of the ratio, $T_{e,fil}\simeq 1.25 T_{e,back}$ is assumed for this case.\\

\begin{figure}
	\centering
		\includegraphics[width=.5\linewidth]{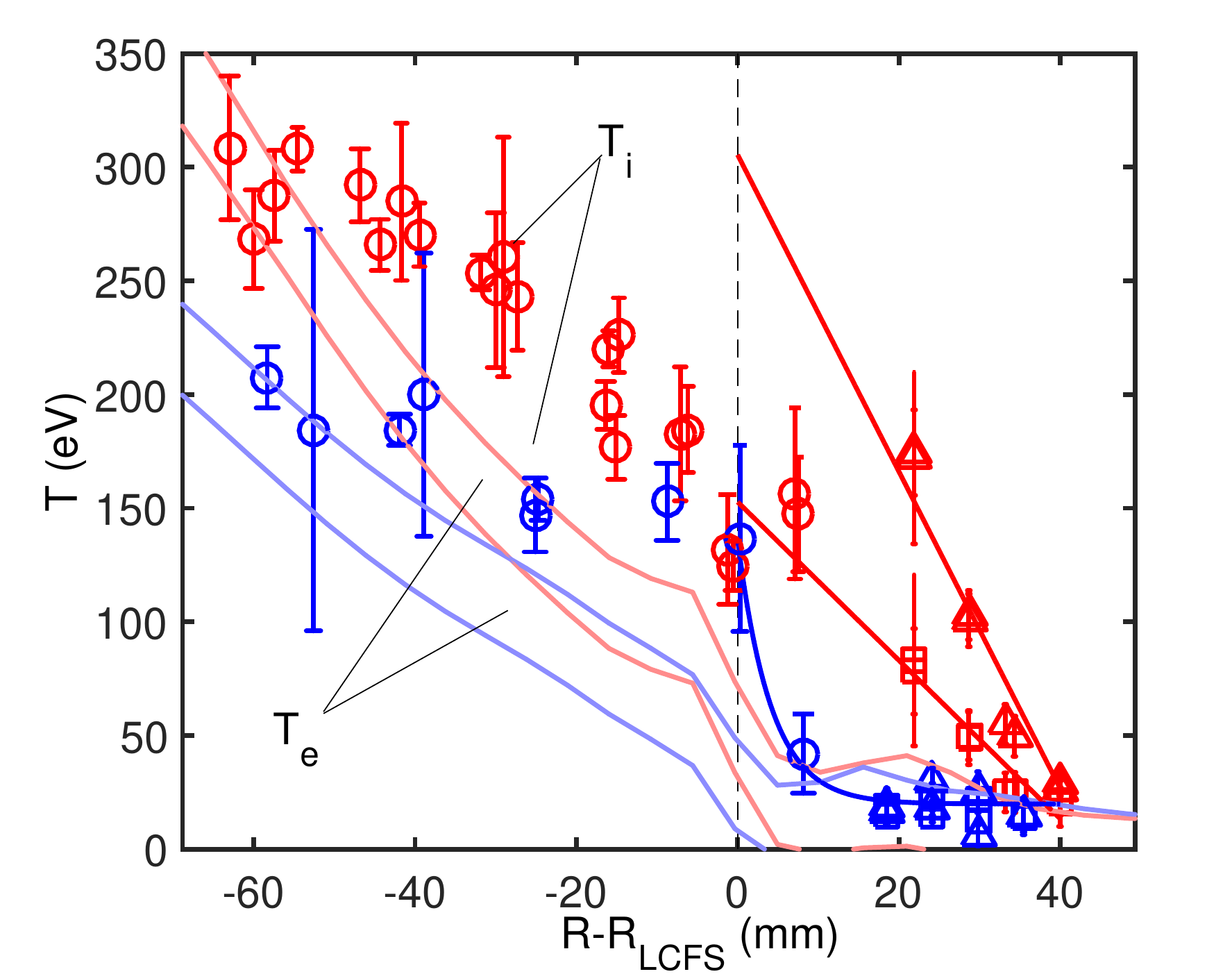}
	\caption{\textit{Edge temperature profiles. Circles indicate CXRS $T_i$ measurements. Triangles/squares represent filament/background $T_i$ data as measured with the RFA probe. Red/blue indicates low/high collisionality. Solid/dashed red lines indicates the fit for background/filaments at low $\Lambda_{div}$. Solid blue curve indicates the fit at high $\Lambda_{div}$. Areas between lines in lighter colors represent IDA $T_e$ profiles, as in Fig. \ref{fig:Te}.}}
	\label{fig:Ti}
\end{figure}

In Fig. \ref{fig:Ti}, $T_i$ measurements for both regimes are plotted. Up to the separatrix, NBI blips were used in order to obtain CXRS data, while the RFA was used to measure $T_{i,fil}$ and $T_{i,back}$ in the far SOL. As has already been discussed elsewhere \cite{Carralero17}, for $\Lambda_{div} \simeq 0.01$, high ion temperatures are observed in the SOL, with a slow radial decay (in the order of $\lambda_{T_i} \simeq 30$ mm) and hot filaments featuring $T_{i,fil} \simeq 2T_{i,back}$. Instead, $T_i$ displays a strong reduction after the shoulder is formed, $T_{i,fil}$ becomes similar to $T_{i,back}$ and the e-folding length drops to $\lambda_{T_i} \simeq 8$ mm  for $\Lambda_{div} \simeq 10$. Similar evolutions have been reported in Alcator C-mod and MAST \cite{Brunner14,Allan16}. The fits used for each case are also displayed in the figure: for the background in the low $\Lambda_{div}$ case, a linear fit is found to be the most appropriated to account for the two sets of data points. As in the case of $T_e$, the average $T_{i,fil}/T_{i,back} \simeq 2$ ratio has been kept for the $T_{i,fil}$ curve. In the high collisionality case, an exponential has been fitted and $T_{i,fil}/T_{i,back} \simeq 1$ has been assumed. The IDA $T_e$ profiles have also been included in Fig. \ref{fig:Ti} in order to compare the temperature profiles of both species. As can be seen, $T_i$ departs from $T_e$ a few cm before reaching the LCFS. This is a well known fact in AUG \cite{Reich04}, but also in other tokamaks such as MAST, Tore-Supra, Alcator C-mod, or JET (see \cite{Kocan11} and references therein), where typically $T_{i,sep} > T_{e,sep}$. In particular, for AUG $T_{i,sep} \simeq 3T_{e,sep}$, regardless of the state of shoulder formation. \\ 

\begin{figure}
	\centering
		\includegraphics[width=.5\linewidth]{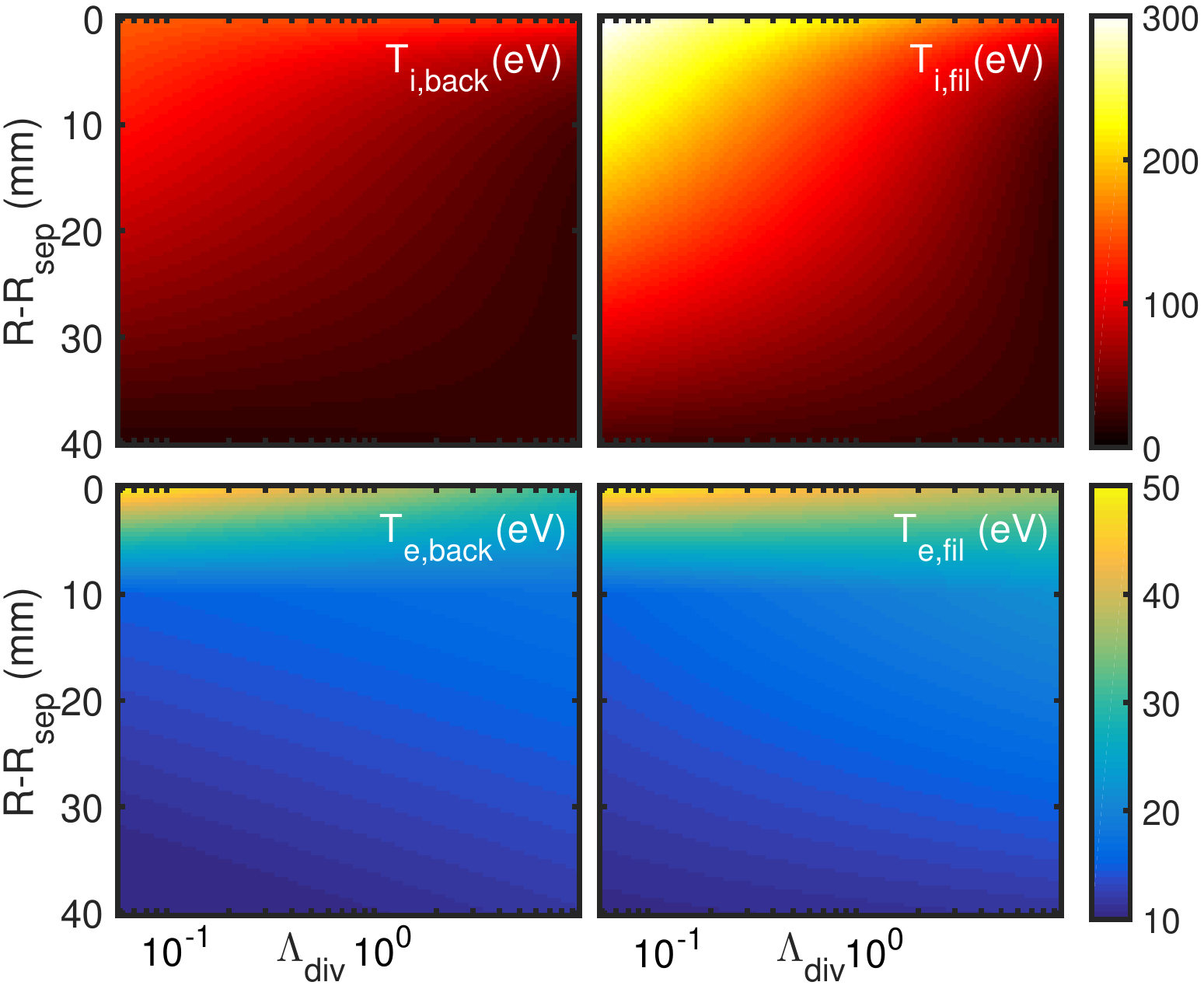}
	\caption{\textit{Interpolated temperature profiles, as a function of $\Lambda_{div}$. Top/bottom plots stand for $T_i$/$T_e$. Left/Right stand for filament/background values. }}
	\label{fig:intTi}
\end{figure}

Finally $T_i$ and $T_e$ profiles were obtained both for filaments and background using $\log(\Lambda_{div})$ to interpolate between the two collisionality values, $\Lambda_{div}^{low} = 0.05$ and $\Lambda_{div}^{high} = 10$, in which the temperature profiles, $T^{\Lambda_{low}}$ and $T^{\Lambda_{high}}$, were measured:

\begin{equation}
T_\alpha=T^{\Lambda_{low}}_\alpha  + \frac{\log({\Lambda_{div}/\Lambda_{div}^{low}})}{\log{(\Lambda_{div}^{high}/\Lambda_{div}^{low}})}(T^{\Lambda_{high}}_\alpha  - T^{\Lambda_{low}}_\alpha ), 
\end{equation}

where $\alpha$ stands for each of the electron/ion, background/filament combinations. The result, displayed in Fig.\ref{fig:intTi}, is a $T_\alpha(R-R_{SOL},\Lambda_{div}$) function for each of the four parameters, which will allow to estimate the SOL temperature for shots in which it was not directly measured. It must be taken into account that, since no filament temperature data is available in the near SOL neither for electrons nor ions, the values assumed there are simply an extrapolation of the trends observed in the far SOL. This will be noted when appropriate during the analysis.\\

\subsection{Divertor}\label{Div}

As explained in section \ref{Exp}, divertor conditions were monitored using the flush mounted triple probes in the target. This system provides direct measurements of $I_{sat}$ and $T_e$, which can be used \cite{Weinlich97} to estimate plasma density, $n_e$, 

\begin{equation}
n_e= 2 \frac{I_{sat}}{Z_iec_sA_{eff}}, \label{eq:ne}
\end{equation}

where $A_{eff}$ is the effective collection area of the probe, $c_s = \sqrt{(T_e+T_i)/m_i}$ is the sound speed, and $Z_i \simeq 1$ can be assumed for the described experiments. As well, using the sheath transmission theory, the perpendicular heat load on the target plates, $q_w$, can be derived from the fixed probes measurements. Indeed, following the development in \cite{Brida17},

\begin{equation}
 q_{w}=\frac{I_{sat}}{A_{eff}Z_ie}(T_e[\gamma_i\tau_i(1-R_E)+\gamma_e-R_E\frac{eV_S}{T_e}]+E_{rec}), \label{eq:qw}
\end{equation}
where $V_S$ is the voltage drop at the sheath, $eV_S/T_e \simeq 2.85$, $E_{rec}$ is the combination of the ionization energy and the Franck–Condon dissociation energy per recycled ion ($E_{rec} = E_{ion} + E_{FC,D_2}/2= 13.6 + \frac{5.5}{2}$ eV), and $\gamma_i$ and $\gamma_e$ are the ion and electron sheath transmission coefficients, $\gamma_i \simeq 2$ and $\gamma_e \simeq 5$. Finally, $R_E$ is the wall energy reflection coefficient which can be calculated using the empirical formula in \cite{Eckstein} for light ion reflection and taking the atomic number and mass of deuterium and tungsten corresponding to the incident and target materials, respectively. This approach is valid for ion impact energies $E_0 < 10^5$ eV, where 
\begin{equation}
E_0 = \gamma_iT_i+eV_S \simeq T_e (\gamma_i\tau_i +2.85),
\end{equation}
meaning that all data in this work falls within such range. When calculating $q_{w}$ in the AUG divertor, $\tau_i = 1$ has additionally been assumed. This is not only consistent with available measurements of $T_i$ in the divertor region \cite{Elmore15}, but also has been proved a right approach for AUG in \cite{Brida17}, where Eq. \ref{eq:qw} was used to calculate heat fluxes onto the targets under several plasma conditions and compare it with independent infrared thermography measurements, showing remarkable agreement between the two.

\begin{figure}
	\centering
		\includegraphics[width=0.7\linewidth]{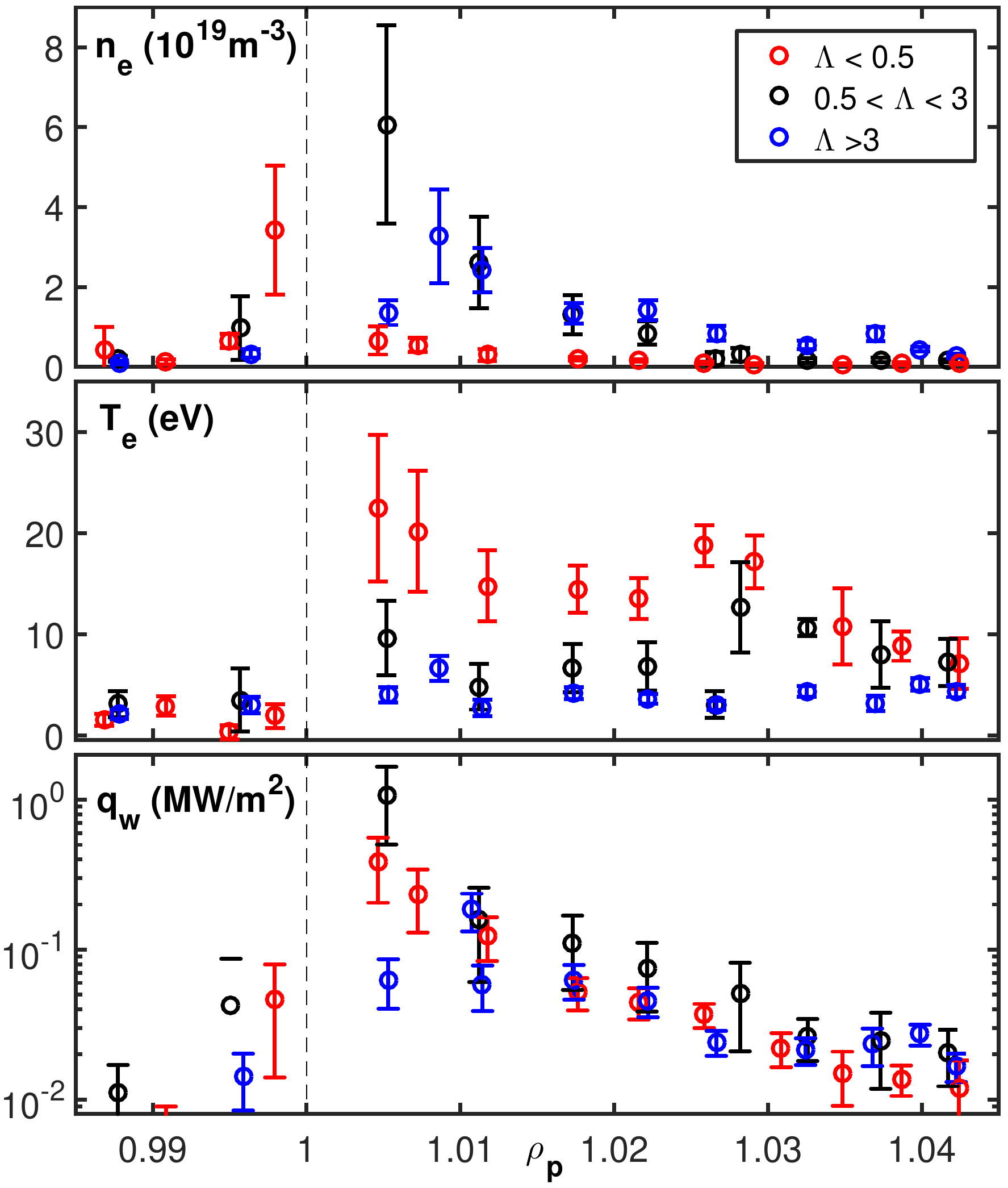}
	\caption{\textit{Average LFS divertor conditions for different ranges of $\Lambda_{div}$. Top/medium/bottom plot shows $N_e$/$T_e$/$q_{w}$ values measured by the fixed Langmuir probes installed in the divertor plates. Low/medium/high $\Lambda_{div}$ values are represented in red/black/blue.}}
	\label{fig:div}
\end{figure}

For each of the points in the database, a divertor profile of $n_e$, $T_e$ and $q_{w}$ has been measured. In Fig. \ref{fig:div}, the average values for each collisionality range are displayed. As discussed in a previous work \cite{Carralero17}, changes in collisionality are associated to the onset of detachment, which transits from an attached state for $\Lambda_{div} \ll 1$ to a partially detached one for $\Lambda_{div} \simeq 5$. This can be seen in the figure: as $\Lambda_{div}$ increases, the points of maximum particle and heat flux move outwards,  $T_e$ drops from around $20$ eV to less than $5$ eV and a clear drop in the power to the wall can be seen in the strike point region, although not in the divertor region corresponding to the far SOL ($\rho_p > 1.01$), where the density remains high enough to sustain $q_{w}$ despite the lower temperature. 

\subsection{Bolometry}\label{Bol}

As explained in section \ref{Exp}, bolometry measurements are used to obtain the amount of power radiated inside the separatrix, $P_{rad}$, in order to calculate $P_{SOL}$. Unfortunately, for the kind of low power, L-mode discharges studied in the present work, the intensity of the radiated power is barely over the detection threshold of the system (this is particularly the case for the low-density plasmas where $\Lambda_{div} \ll 1$). As well, most radiation comes from the X-point region, where the distinction between radiation originated inside or outside the LCFS becomes particularly difficult. As a result, no detailed characterization of $P_{SOL}$ as a function of $\Lambda_{div}$, input power, density, etc., was possible. Instead, based on the available data for this set of shots, it was roughly estimated that $P_{rad} \simeq 0.15 P_{tot}$ for $\Lambda_{div} < 1$ and $P_{rad} \simeq 0.25 P_{tot}$ for $\Lambda_{div} > 1$, where $P_{tot} = P_{ECH}+P_{ohm}$ is the total input power.

\section{Database validation}\label{sec:validation}

Once the database from probe measurements has been compiled, the interpolations can be validated with independent diagnostics. This is done in two ways: first, midplane density profiles are calculated and compared to LiB data. Second, the heat flux into the MPM probe head is calculated based on local plasma parameters and compared with the values obtained from thermography.

\subsection{Density}\label{Den}

Once $T_e$ and $T_i$ are defined for the different $\Lambda_{div}$ and the whole $R-R_{sep}$ range, it is now possible to obtain density values from MPM $I_{sat}$ measurements using eq. \ref{eq:ne}. In this case, $A_{eff} = 3.2$ mm$^2$ is used, corresponding to the effective pin surface. Also, for each point, the $I_{sat}$, $T_{e}$ and $T_{i}$ values associated to filaments and background have been used separately in order to produce the corresponding $n_e$ values.

\begin{figure}
	\centering
		\includegraphics[width=0.5\linewidth]{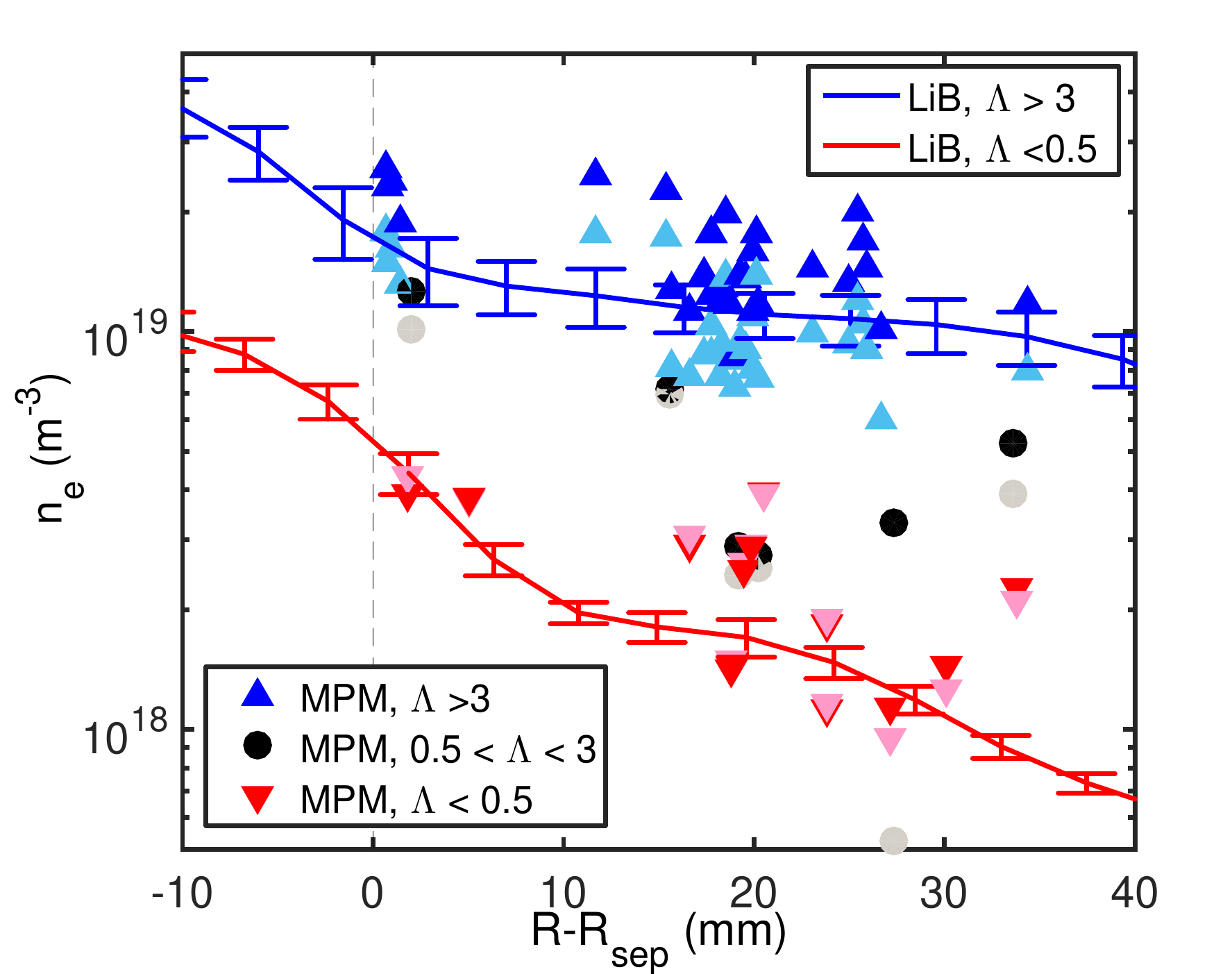} 
	\caption{\textit{Comparison of upstream radial density profiles. Symbols stand for probe measurements, with normal/light shades of the color code used to indicate filament/background values. Red/blue solid lines are typical LiB profiles for low/high $\Lambda_{div}$ values.}}
	\label{fig:ne}
\end{figure}

The resulting density profiles are displayed in Fig. \ref{fig:ne} for both filaments and background, with the latter in the same color/marker scheme as usual, but with lighter tones of red/black/blue. For comparison, density profiles from the lithium beam diagnostic corresponding to high and low $\Lambda_{div}$ values are displayed as solid lines. As can be seen, a very good agreement exists between the two diagnostics: the high/low density curves from the LiB fall between the background and filament density levels measured by the probe for the corresponding $\Lambda_{div}$ discharges. A clear increase for both filament and background density can be seen when $\Lambda_{div} \simeq 1$ is surpassed, indicating the shoulder formation. This result is properly reproduced by data points measured at high collisionality discharges. It is interesting to note that the fact that $T_{e,fil}$ and $T_{i,fil}$ are extrapolated in the near SOL does not seem to affect strongly the comparison with the LiB data. Although the influence of the temperature  in this measurement is reduced ($n_e \propto I_{sat}(T_e+T_i)^{-1/2}$), this indicates that the extrapolation is at least qualitatively correct. \\

\subsection{Local heat flux}\label{lhf}

\begin{figure}
	\centering
		\includegraphics[width=0.5\linewidth]{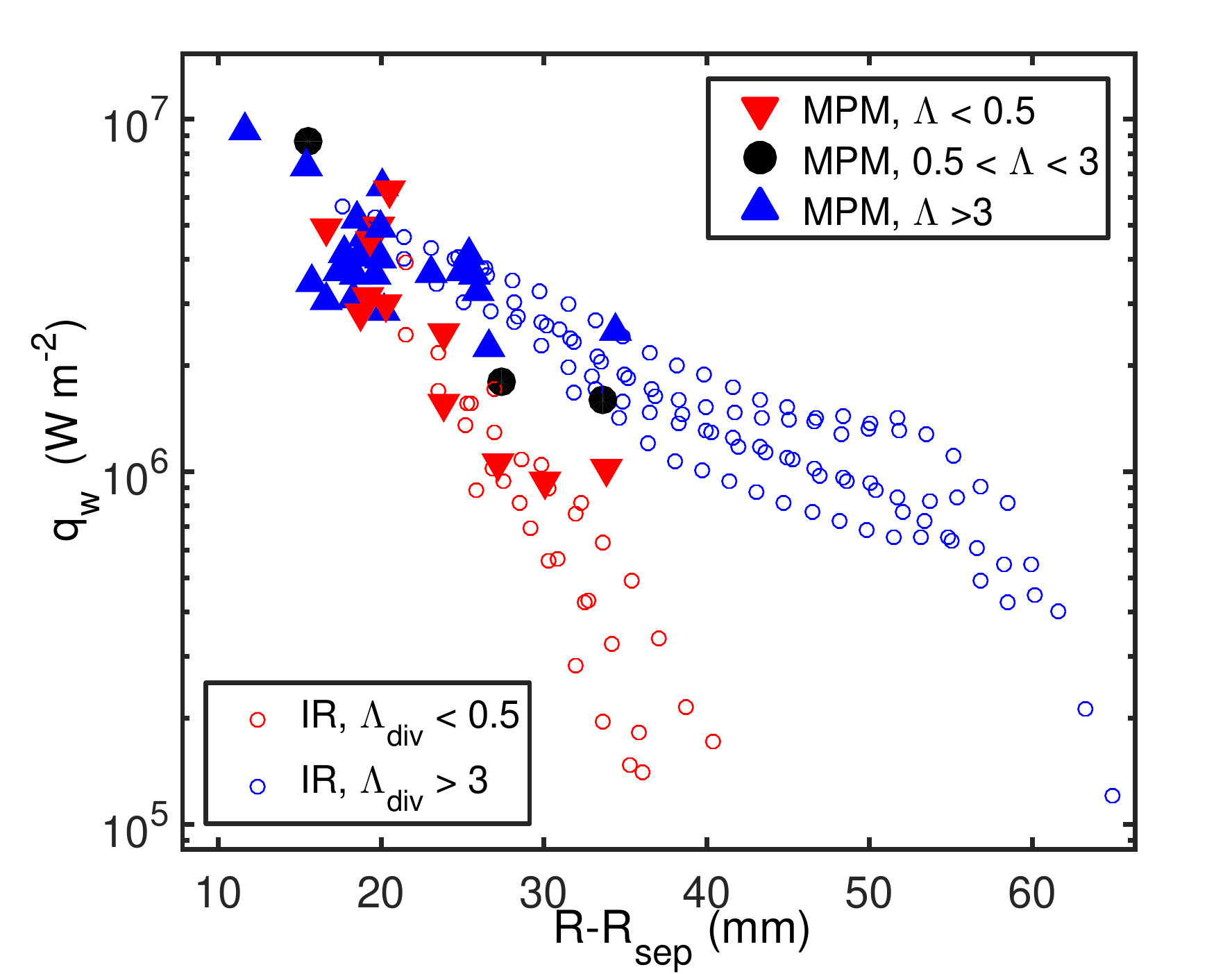} 
	\caption{\textit{Heat flux onto the MPM. Open circles represent THEODOR code calculations of $q_{w}$ carried out from IR measurements. Closed circles and triangles represent $q_{w}$ calculated using the sheath transmission Eq. \ref{eq:sheath_trans} and probe data. Color code as in Fig. \ref{fig:Turb}. }}
	\label{fig:qwMPM}
\end{figure}

The heat flux deposited by the plasma onto the MPM probe head can be calculated by adapting the method discussed in section \ref{Div} for the divertor to the midplane conditions: filament and background contributions, which could not be separated in the divertor, must now be combined in order to obtain a total heat flux equivalent to the one derived from the IR data. In order to do this, the following assumption is made: for a given period of time $t_{sample}$, turbulence properties in the SOL, such as density, temperature, etc. correspond to those of filaments for $f_{fil}t_{sample}$, and to those of the background for $(1-f_{fil})t_{sample}$. Therefore, the heat flux onto the wall associated to filaments, $q_{w}^{fil}$, and to the background, $q_{w}^{back}$, are calculated using Eq. \ref{eq:qw} but taking the corresponding  $I_{sat}$, $T_{e}$ and $T_{i}$ values from the database (meaning that $\tau_i = 1$ generally no longer applies), and the total heat flux onto the probe wall is calculated as

\begin{equation}
 q_{w}= f_{fil} q_{w}^{fil} + (1-f_{fil}) q_{w}^{back}.  \label{eq:sheath_trans}
\end{equation}
The results are displayed in Fig. \ref{fig:qwMPM}: two clear tendencies can be observed before and after the transition: starting from a common value of  $ q_{w} \simeq 5$ MWm$^{-2}$ at $R-R_{sep} \simeq 20$ mm, a much slower decay is observed after the shoulder formation, with intermediate collisionality points falling nicely between the two other. This observation is in remarkable agreement with the results obtained from the IR camera, $q_{w}^{IR}$: In Fig. \ref{fig:qwMPM}, the $q_{w}^{IR}$ values calculated at the radial strip highlighted in red in Fig. \ref{fig:IR} are  also presented. The color code indicates the $\Lambda_{div}$ value at the time of the corresponding plunge and also indicates a clear change in the parallel heat flux with the shoulder formation: In good agreement with previous work \cite{Carralero14}, the e-folding length at the probe wall, $\lambda_q$ goes from $ \simeq 5$ mm up to $\simeq 18$ mm over the transition. In this case, since no IR data is available in the near SOL for comparison, the fact that $T_{e,fil}$ and $T_{i,fil}$ have been extrapolated (see Section \ref{Temp}) does not have any impact.

\section{Perpendicular transport associated to filaments}\label{sec:transport}

After the probe database has been validated, it can be finally used to calculate the perpendicular transport associated to filaments in the far SOL.

\subsection{Perpendicular particle transport}

\begin{figure}
	\centering
		\includegraphics[width=0.5\linewidth]{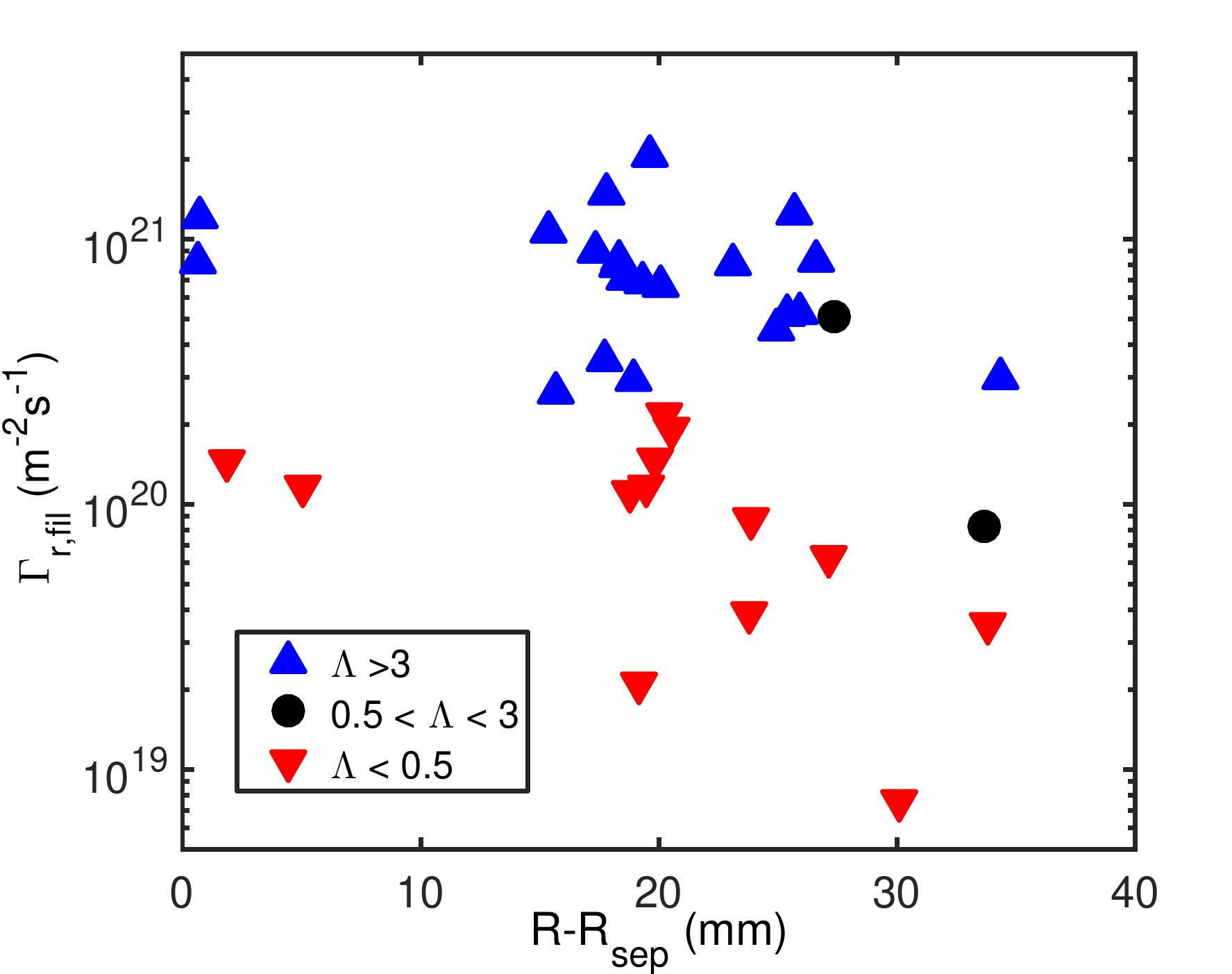}
	  \includegraphics[width=0.5\linewidth]{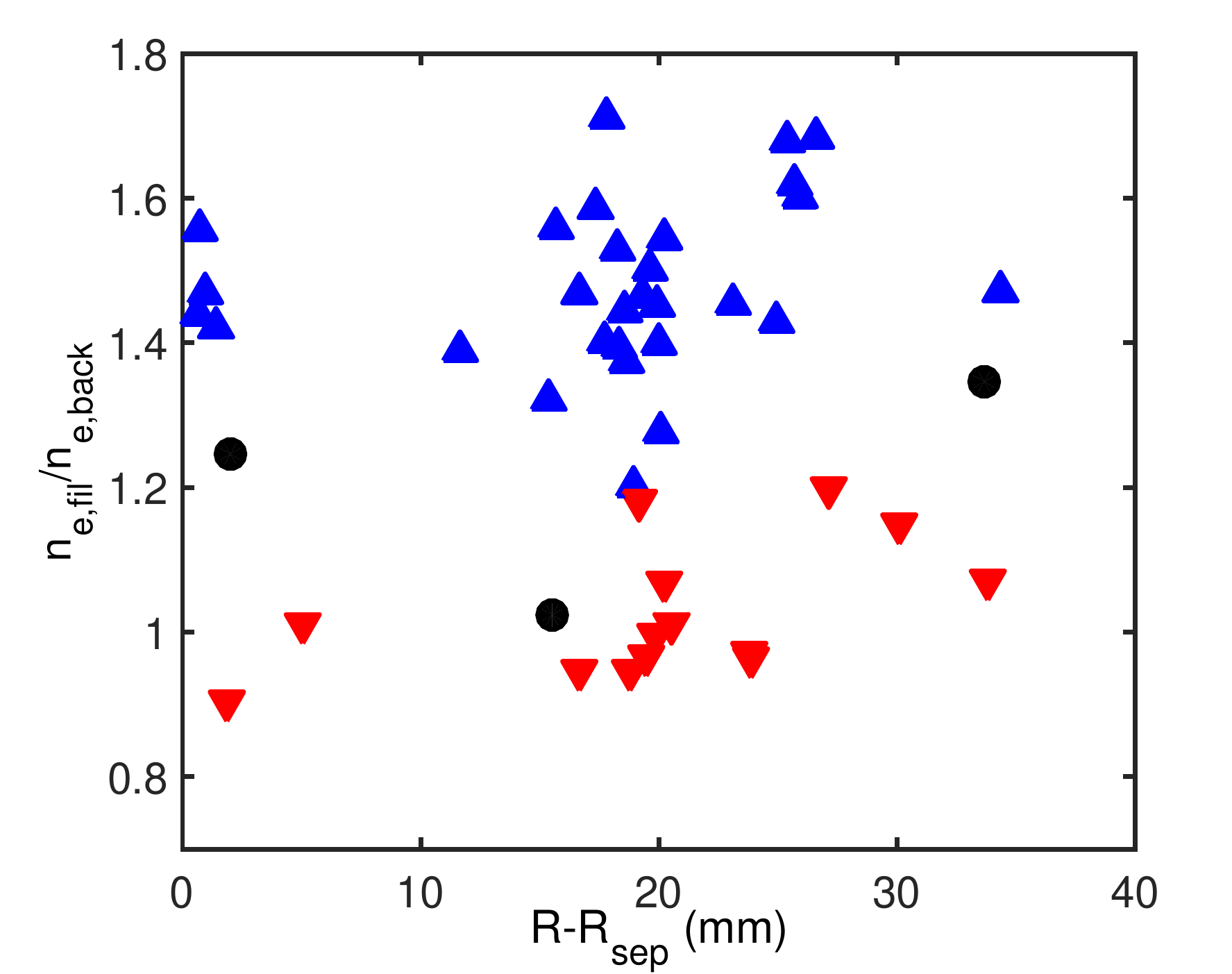} 
	\caption{\textit{Particle flux associated to filaments. Top) $\Gamma_{r,fil}$.  Marker/color code as in Fig. \ref{fig:Turb}. Bottom) Filament to background density ratio.  Both are presented as a function of $R-R_{sep}$.}}
	\label{fig:G}
\end{figure}

First, the particle transport associated to filaments can be obtained from the magnitudes discussed in section \ref{Mes}, as the number of particles traveling radially in an average filament, factored by the fraction of the time in which a filament is present at a given point of the far SOL:

\begin{equation}
 \Gamma_{r,fil}=v_rn_{fil}f_{fil} \label{eq:G}
\end{equation}
The results are displayed in Fig. \ref{fig:G}, where the values of $\Gamma_{r,fil}$ are displayed as a function of $R-R_{sep}$. As can be seen, two clear profiles emerge for the low and high collisionality cases, the second almost one order of magnitude higher than the first. It must be pointed out that this is not just the trivial consequence of the density increase associated to the shoulder formation (which can be seen in Fig. \ref{fig:ne}): as can be seen in the bottom plot of Fig. \ref{fig:G}, where the profile of the $n_{e,fil}/n_{e,back}$ ratio is presented, the amplitude of density fluctuations is also greatly increased after the transition. This is in good agreement with equivalent measurements carried out over the shoulder formation at TCV \cite{Garcia07}.

It must be noted that this definition of $\Gamma_{r,fil}$ only accounts for the transport caused by large convective structures (at least large enough to be detected using the $2.5\sigma$ threshold), and neglects diffusive transport and convective transport associated to smaller structures. In principle, this seems to be a reasonable approximation as the general consensus in literature \cite{Dippolito11} points in the direction of far SOL transport being dominated by large convective cells. Besides, density profiles flatten for high collisionalities (when the shoulder is formed), thus reducing the relevance of the diffusive transport. However, a detailed comparison of diffusive vs. convective perpendicular transport is left for future work. 

\begin{figure}
	\centering
		\includegraphics[width=.5\linewidth]{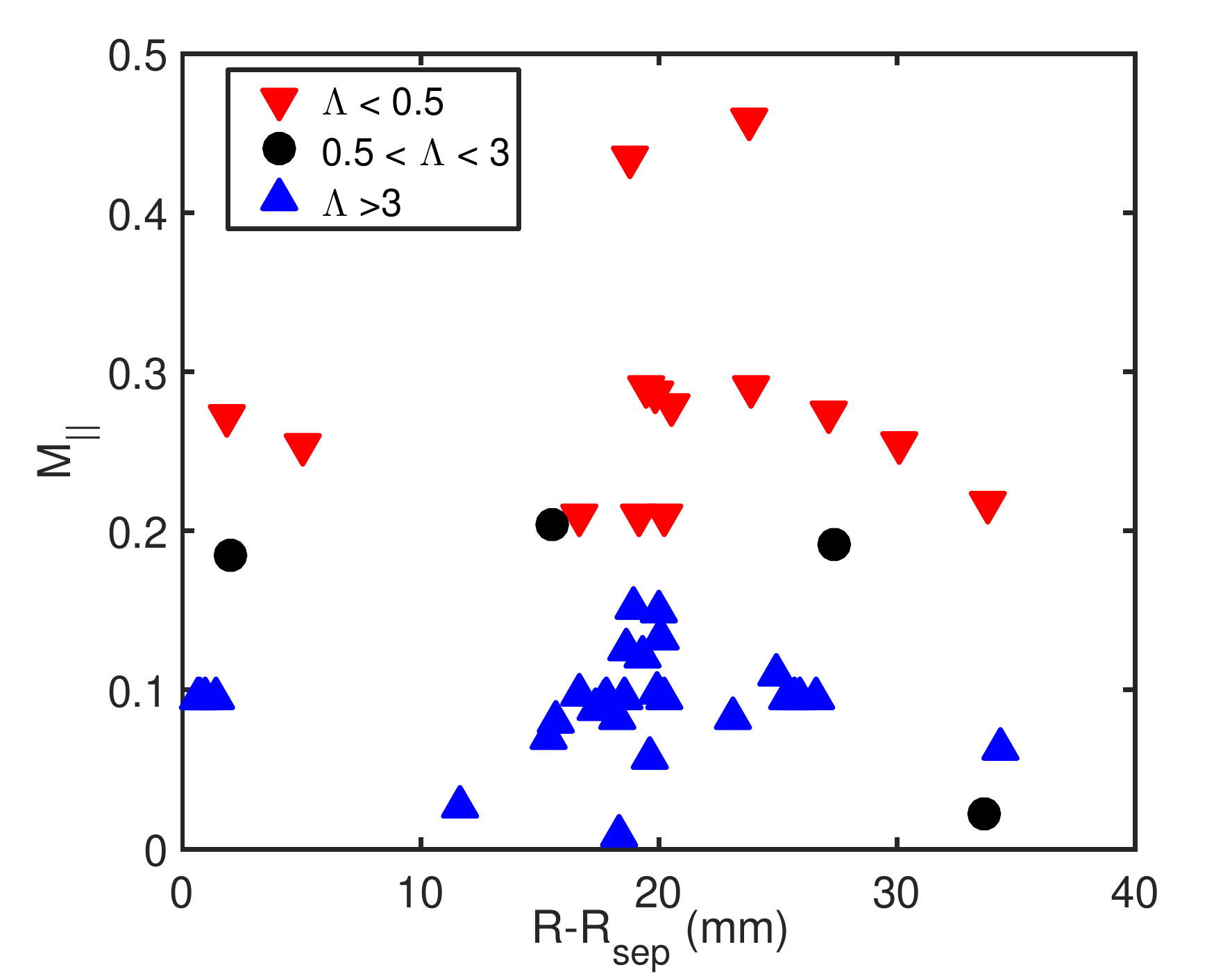}
		\includegraphics[width=.5\linewidth]{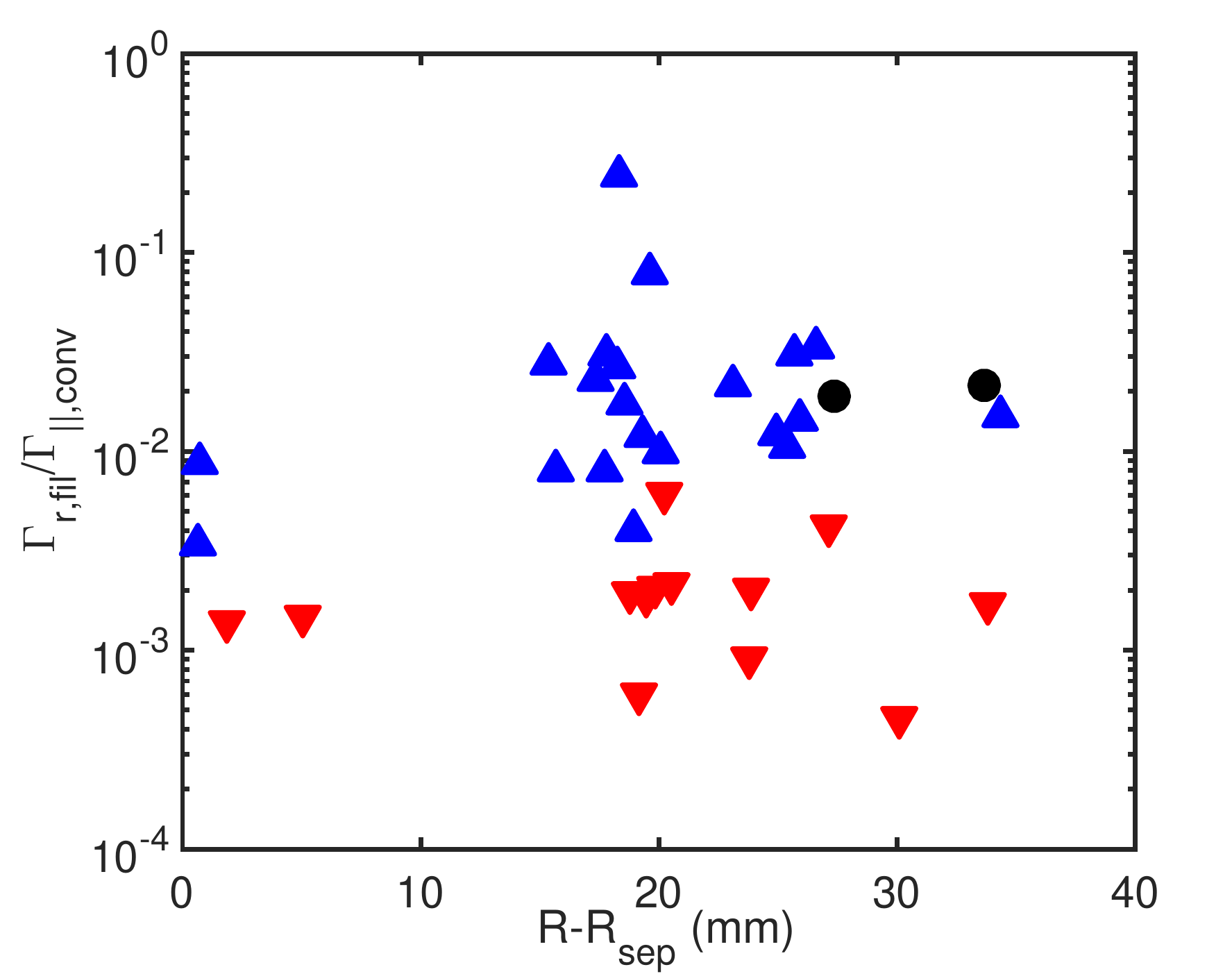}
	\caption{\textit{ Mach number measurements. Top) Mach number profile. $M_{||} > 0$ indicates a flow towards the HFS divertor. Bottom) Perpendicular to parallel transport ratio profile. Same color code as in Fig. \ref{fig:Turb}.}}
	\label{fig:Mach}
\end{figure}

In order to discuss changes in the perpendicular to parallel transport balance, measurements from the Mach probe are presented in Fig. \ref{fig:Mach}: In it, the radial profile of the Mach number in the direction parallel to the magnetic field, $M_{||}$, at the probe position in the midplane is displayed for the different collisionality regimes ($M_{||} > 0$ indicates a flow towards the HFS divertor). The low collisionality regime features higher values, in the range of $M_{||} \simeq 0.3$. Instead, for $\Lambda_{div} > 3$ $M_{||}$ drops roughly by a factor of three, with intermediate collisionalities sitting nicely between the other two clouds of points. This may be seen as the result of a transition between a sheath-limited divertor regime for low $\Lambda_{div}$, in which the main particle source in the SOL is still the separatrix, to a high recycling and detached regime for high collisionalities, in which recycling in front of the target becomes the main particle source in the SOL, thus drastically reducing the parallel velocity upstream. Again, if the filamentary transport is assumed to dominate perpendicular particle transport, these measurements allow for the comparison between the parallel and perpendicular convection terms,

\begin{equation}
\frac{ \Gamma_{r,fil}}{\Gamma_{||,conv}}=\frac{n_{e,fil}}{n_{e,mean}}\frac{f_{fil}v_r}{c_{s,mean}M_{||}},
\end{equation}
where $n_{e,mean}= f_{fil} n_{e,fil} + (1-f_{fil}) n_{e,back}$ and $c_{s,mean}= f_{fil} c_{s,fil} + (1-f_{fil}) c_{s,back}$, since only an average $M_{||}$ is measured. This is displayed in the bottom plot of Fig. \ref{fig:Mach}, where it can be seen how, even if the parallel term dominates in both cases, a clear change is observed in the far SOL, where the ratio  increases by almost an order of magnitude for $\Lambda_{div} >3$. This is consistent with the hypothesis that the filamentary transition plays a role in the formation of the shoulder, as the perpendicular to parallel particle balance is clearly changed for high collisionalities. Since $n_{e,fil}$ includes a $(T_e+T_i)^{-1/2}$ factor, the set of the points in the near SOL could be affected by the extrapolation of $T_i$ values. Nevertheless, they seem to follow the same trend as the rest of points and $n_{e,fil}$ values seem qualitatively correct in Fig. \ref{fig:ne}, so probably the picture in Fig. \ref{fig:Mach} is at least roughly correct.

\subsection{Perpendicular heat flux}

\begin{figure}
	\centering
		\includegraphics[width=0.5\linewidth]{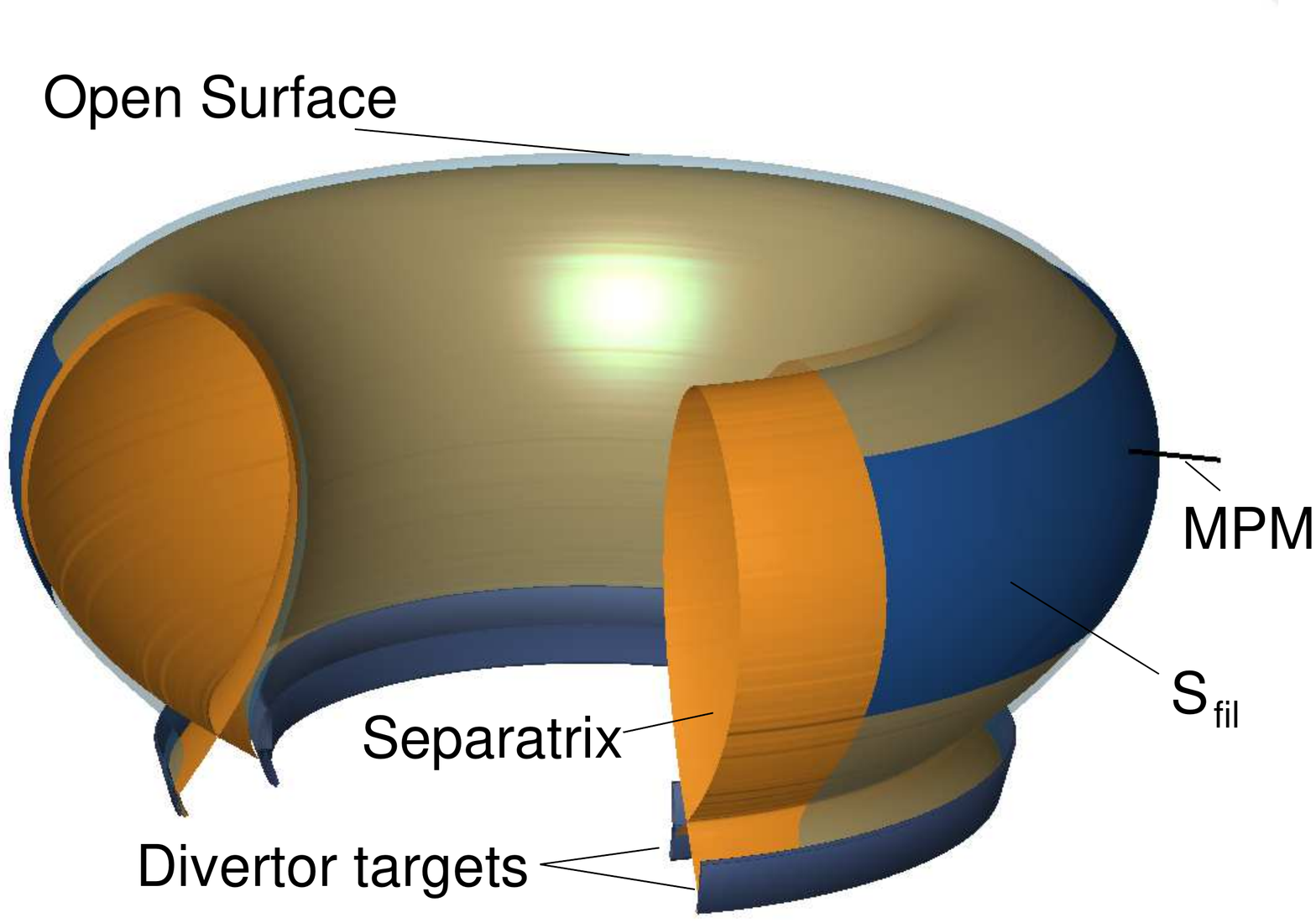}
		\includegraphics[width=0.48\linewidth]{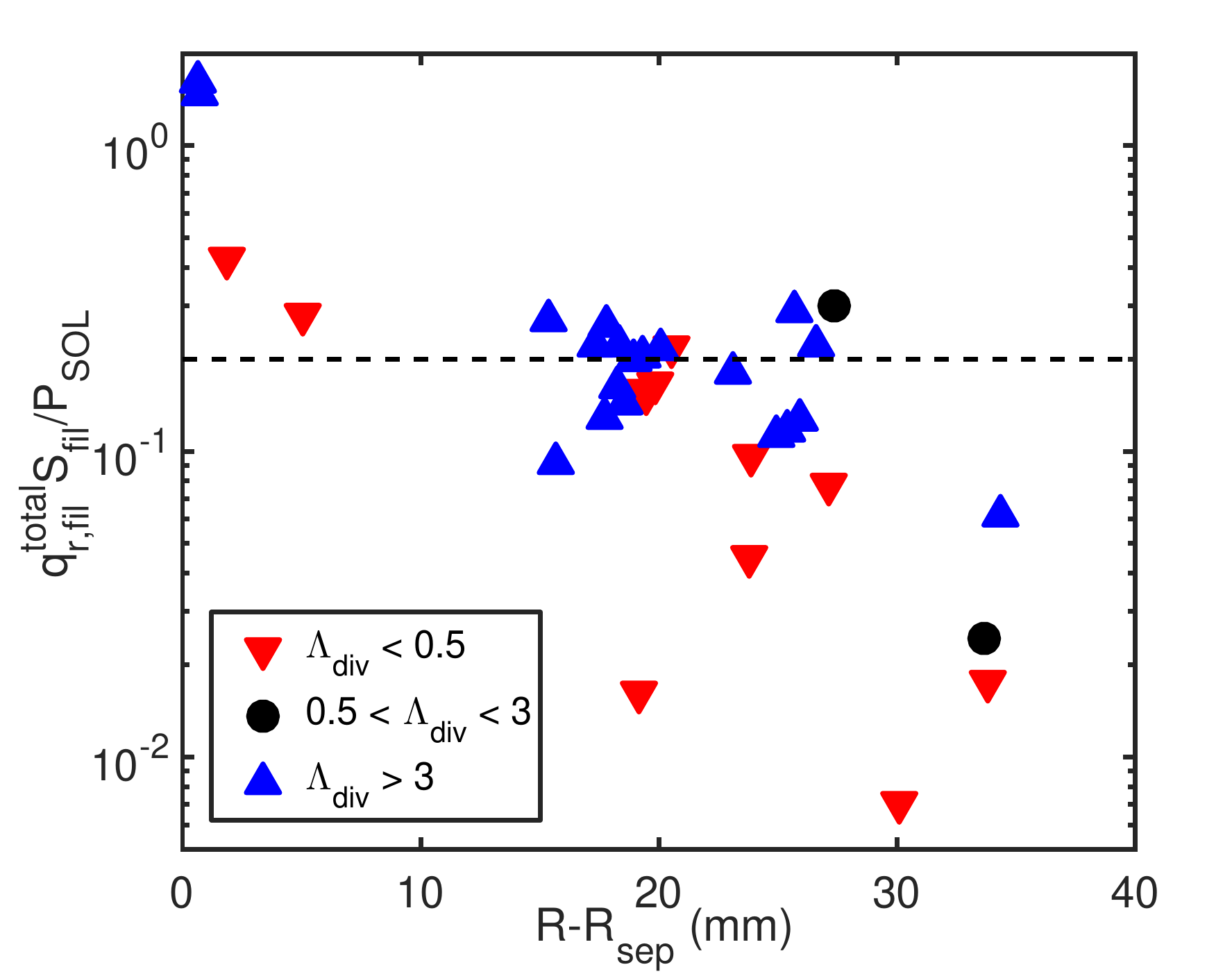}
	\caption{\textit{Global heat transport associated to filaments. Left) Schematic representation of the LCFS, the open flux surface associated to a given probe position and the corresponding surface, $S_{fil}$, defined as in Eq. \ref{eq:Sfil}. Right) Profiles of the estimated total perpendicular power flux, $q_{per}S_{fil}$, normalized to the total power through the separatrix, $P_{SOL}$, for low/medium/high collisionality (in red/black/blue symbols).}}
		\label{fig:P}
\end{figure}

Finally, once $\Gamma_{r,fil}$ is known from Eq. \ref{eq:G}, temperature measurements can be used to calculate the radial heat flux associated to filaments \cite{Stangeby},

\begin{equation}
 q_{r,fil}^{total}=q_{r,fil}^{electron}+q_{r,fil}^{ion}=\frac{5}{2}\Gamma_{r,fil}T_{e,fil}+\frac{1}{2}\Gamma_{r,fil}[5T_{i,fil}+m_i(c_sM_{||})^2], 
\end{equation}
where $m_e \ll m_i$ has been taken into account. In order to compare this flux with the total power flowing into the SOL, $P_{SOL}$, the total power crossing the corresponding flux surface can be defined as

\begin{equation}
P_{fil}=q_{r,fil}^{total}S_{fil}, \label{eq:Pfil}
\end{equation}
where a characteristic surface $S_{fil}$ is defined for each data point. This surface, displayed in Fig.\ref{fig:P}, represents the region of the flux surface where the MPM measurement is made, in which there is filamentary transport. Given the ballooning nature of the filaments, this is only expected to happen where the curvature is unfavorable \cite{Terry03,Antar05,Asakura09}, meaning that only a poloidal region around the low field side midplane should be considered. Considering the axial symmetry of a tokamak, $S_{fil}$ can then be defined in cylindrical coordinates as 

\begin{equation}
S_{fil} = \int_{z(-\theta_{max})}^{z(\theta_{max})}{2\pi R(z)dz}, \label{eq:Sfil}
\end{equation}
where $\theta$ is the poloidal coordinate (with $\theta = 0$ indicating the outer midplane), $R(\theta)$ is the radial coordinate of the corresponding flux surface and $\theta_{max}$ is the poloidal angle limiting the region with filamentary transport. In order to calculate $R(\theta)$, a field line is traced from the MPM measurement point to the LFS divertor, keeping track of its $z(\theta)$ and $R(\theta)$ components. In order to decide the extent of $S_{fil}$, $\theta_{max} \simeq \pi/3$ has been assumed, resulting in typical values of $S_{fil}$ in the range of $13$ m$^2$. This must be regarded as an approximation: a constant value of $q_{r,fil}^{total}$ has been assumed in eq. \ref{eq:Pfil}, corresponding to the one measured at the probe position (roughly $\theta_{MPM} \simeq \pi/6$). In fact, the filamentary drive (and thus $\Gamma_{r,fil}$) can be expected to increase closer to the equator and decrease further away. To test this definition of $P_{fil}$, the simple hypothesis $q_{r,fil}(\theta)=q_{r,fil}^{eq}\cos{\theta}$ can be made, where $q_{r,fil}^{eq}$ is the radial heat flux at the equator. In order to match the MPM measurements, $q_{r,fil}^{eq}=q_{r,fil}^{total}/\cos{(\theta_{MPM})}$. In this case, equation \ref{eq:Pfil} reads now $P_{fil}= q_{r,fil}^{total}S'_{fil}$, where
\begin{equation}
S'_{fil} = \int_{z(-\theta_{max})}^{z(\theta_{max})}{\frac{2 \pi}{\cos{(\theta_{MPM})}} \biggl(1+(z-z_{mag})^2/(R(z)-R_{mag})^2\biggr)^{-1/2} R(z) dz} , \label{eq:Pfil2}
\end{equation}

and $z_{mag}$ and $R_{mag}$ are the coordinates of the magnetic axis. By integrating the new effective area, $S'_{fil}$, it is observed that the total radial heat flux under this poloidal distribution remains unchanged close to the separatrix and decreases slightly (up to $5\%$) for the data points with higher $R-R_{sep}$ values. Of course, this result depends on the specific function selected as $q_{r,fil}(\theta)$, but as a first approximation, the error introduced by the poloidally constant flux hypothesis is expected to be small and a detailed discussion of this problem is left for future work.\\

In Fig. \ref{fig:P}, the $P_{fil}/P_{SOL}$ ratio is displayed. As can be seen, the perpendicular power flux associated to filaments may be a substantial fraction of $P_{SOL}$ for radial positions in the near SOL (ideally reaching $P_{fil}/P_{SOL} = 1$ at $R-R_{sep} = 0$ mm). Nevertheless, given the extrapolation involved in the filament temperatures these data points must be regarded with caution. In particular, the fact that high collisionality points near the separatrix show $P_{fil}/P_{SOL} > 1$ can be regarded as a overestimation of $T_{i,fil}$ and/or $T_{e,fil}$, but also as an overestimation of $P_{rad}$ for those particular cases. In the far SOL, there seems to be a radial position around $R-R_{sep} = 20$ mm where the heat fluxes are equal for low and high collisionalities. Beyond this point, both regimes decay at different rates, with the higher collisionality retaining higher values for equivalent radial positions. These results are reminiscent of those displayed in Fig. \ref{fig:qwMPM}. This is not a trivial result, as the evolution of $\Gamma_{r,fil}$ was not involved in the calculation of heat deposition on the MPM surface. As a general result, it can be seen how the flows associated to filaments still represent a significant fraction of $P_{SOL}$ 20 mm in front of the separatrix (and beyond 25 mm in the medium and high collisionality cases).\\

It must be taken into account that $S_{fil}$ is roughly the same for all data points. The same holds true for $P_{SOL}$, as the input power $P_{ohm}+P_{ECH}$ is approximately equal for all shots and $P_{rad}$ does not change much with the transition, as discussed in section \ref{Bol}. This means that the values of $P_{fil}/P_{SOL}$ displayed in Fig. \ref{fig:P} are basically proportional to the ones of $q_{r,fil}^{total}$. \\  

\begin{figure}
	\centering
		\includegraphics[width=0.5\linewidth]{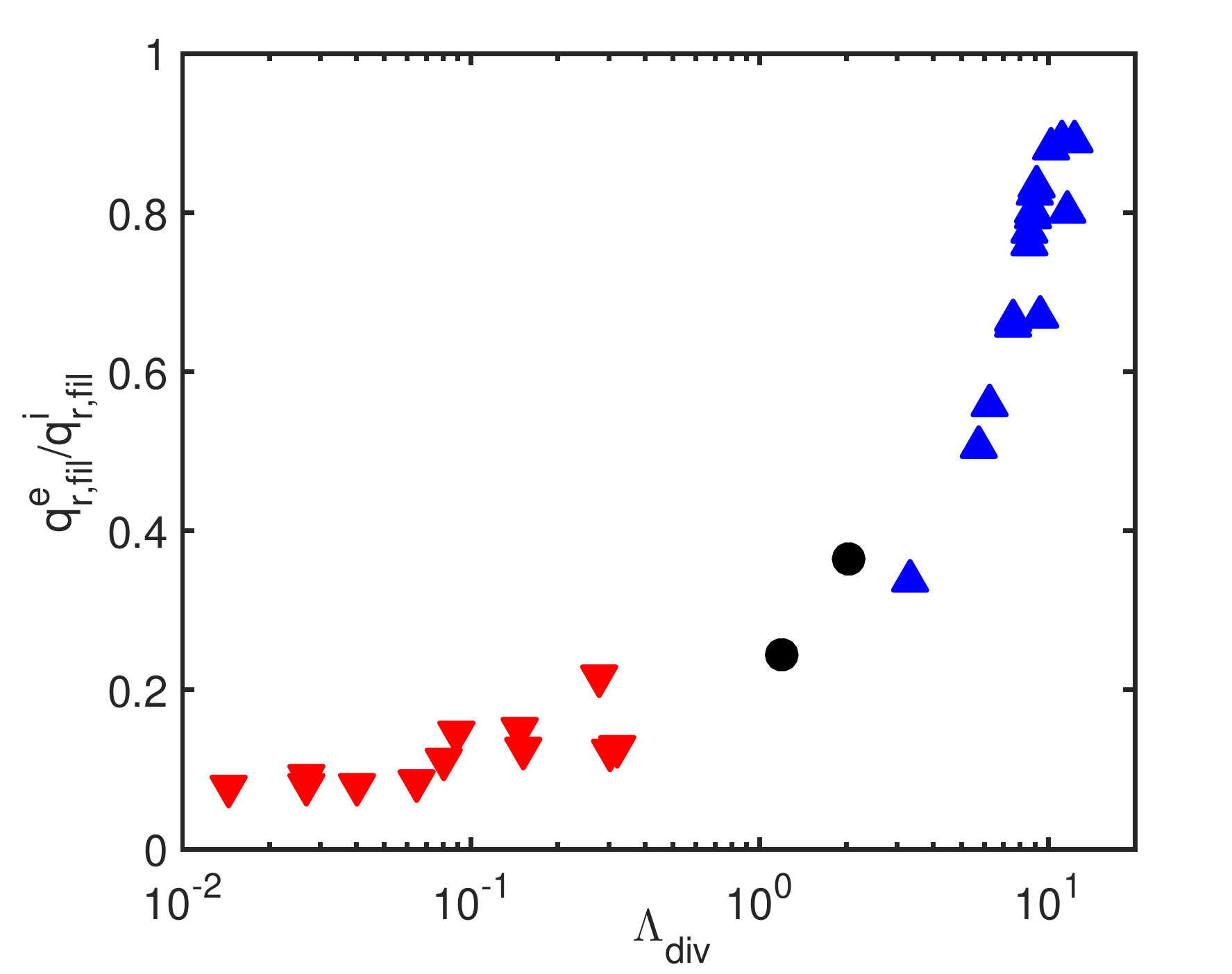}
			\caption{\textit{Ratio of electron to ion perpendicular heat flux associated to filaments as a function of divertor collisionality.}}
		\label{fig:qeqi}
\end{figure}

Finally, the different contributions to the perpendicular heat flow of ions and electrons are displayed in Fig. \ref{fig:qeqi}. As can be seen, a clear evolution takes place for all radial positions: at low collisionalities, $T_i \gg T_e$, meaning that the perpendicular energy transport is dominated by the ion channel. Instead, as the transition takes place ions cool down, and their relative contribution to transport decreases. Eventually, both components reach similar values as $T_i \simeq T_e$ for $\Lambda_{div} \gg 1$.\\

\section{Discussion and conclusions}\label{Con}

In this work, we set out to find out whether the density increase in the SOL caused by the shoulder formation is followed by an equivalent increase of energy transport. By analyzing  data from multiple diagnostics on AUG, we find some relevant answers to this question. First, a quantitative calculation of the particle and heat fluxes associated to filaments is done by using different probes in equivalent L-mode discharges. These data are consistent with previous studies of the shoulder formation carried out in AUG \cite{Carralero14,Carralero17}: the shoulder forms after a certain critical value of divertor collisionality $\Lambda_{div} \simeq 1$ is achieved, filaments become disconnected and are no longer well described by the sheath-limited scaling, and ion temperatures are greatly reduced across the whole SOL after the transition. This approach seems validated by the good agreement found between plasma parameters calculated by combined probe data and independent diagnostics. In particular, the comparison between the heat flux onto the MPM surface calculated using sheath transmission formulas and the one obtained from infrared camera data using the THEODOR code not only shows a remarkable agreement, but also indicates a change in the far SOL parallel heat flux e-folding length from $\lambda_q \simeq 5$ mm before the shoulder formation to $\lambda_q \simeq 18$ mm afterwards.\\

The radial convective particle flux associated to filaments (ie., not including diffusion of the background plasma) is found to increase after the shoulder formation by almost an order of magnitude. This can be partly attributed to the general increase of the SOL density, but also to an enhancement of the filament amplitude and packing fraction. This fact, combined with a drop in the parallel Mach number as measured in the midplane, leads to substantial increase in the radial to parallel particle convective transport ratio, in particular in the far SOL. Although parallel convection still remains dominant, this effect would be consistent with the filamentary transition playing a role in the formation of the shoulder. Regarding radial heat flux, temperature measurements associated to filaments in front of the separatrix are not reliable enough for a quantitative analysis. However, in the far SOL the picture is in good agreement with IR data: $20$ mm in front of the separatrix, radial transport is the same before and after the shoulder formation, and represents around $20\%$ of the power crossing the separatrix. This is the result of the drop in $T_i$ canceling out the increase in $\Gamma_r$, as can be deduced from the fact that, for $\Lambda_{div} \ll 1$, the ion channel dominates $q_{r,fil}$, while for $\Lambda_{div} \gg 1$ ion and electron channels roughly equilibrate. Instead, for $R-R_{sep} > 20$ mm, the radial decay of $q_{r,fil}$ is much slower after the transition. \\

These results provide the first systematic empirical evaluation of the perpendicular heat transport in a tokamak measured directly in the SOL over several transport regimes. The main implication of these results is that a substantial amount of the energy ejected through the separatrix (up to one fifth of $P_{SOL}$) would remain in the midplane $20-25$ mm in front of the separatrix, either by the convection of hot ions before the transition, or simply by the enhanced perpendicular transport afterwards. This leads to two main conclusions which may be important for operation and design of next generation fusion devices, such as ITER and DEMO:\\

First, not all power is deposited on the divertor targets after a few mm, as is commonly assumed in simple SOL descriptions (such as the basic 2 Point Model, as described in sections 5.2 and 5.4 of \cite{Stangeby}). This is particularly true if a large fraction of the power into the separatrix is carried by the ions, thus invalidating the common assumption $T_i \simeq T_e$, according to which parallel transport is dominated by Spitzer-Harm electron conduction. Instead, if as seen in Fig. \ref{fig:Ti}, $T_i \simeq 3 T_e$, the parallel conduction ratio $q_{\parallel,e}/q_{\parallel,i} \simeq \sqrt{m_i/m_e}(T_e/T_i)^{7/2}$ becomes $\simeq 1$. A proper discussion of the implications of this result would require a detailed calculation of the parallel transport and a complete calorimetry study in the SOL of AUG, which are out of the scope of this work and will be addressed in a forthcoming study.\\

A second conclusion would be that, since convection can carry a substantial fraction of $P_{SOL}$ tens of mm away from the separatrix, SOL transport associated to filaments may have a relevant impact on the distribution of heat fluxes onto PFCs. As discussed elsewhere \cite{Carralero17}, this effect could be enhanced in future machines, in which the increase of divertor collisionality leading to an enhanced $\Gamma_r$ might not be immediately correlated to a decrease in the SOL temperature, as seems to be the case in present day machines. This may lead to beneficial effects, such as a partial spread of the loads at the divertor targets, but also to detrimental ones, such as potentially dangerous heat loads onto the main wall PFCs.\\

Further conclusions can't be reached from the data presented in this work, as all of them have been collected in L-mode discharges, which is not the relevant scenario for next-generation plasmas. However, ELMs decay into filaments which may propagate radially by similar physics mechanisms as the ones discussed in this work. The extension of this analysis to H-mode plasmas, and the discussion of shoulder formation and the contribution of filamentary transport in an ELMy SOL is thus left as a future line of work.\\

\section*{Acknowledgements}

This work has been carried out within the framework of the EUROfusion Consortium and has received funding from the Euratom research and training programme 2014–2018 under grant agreement No 633053. The views and opinions expressed herein do not necessarily reflect those of the European Commission.\\

\end{document}